\documentclass{aa}
\bibpunct{(}{)}{;}{a}{}{,} 
\usepackage{graphicx}
\usepackage{color}
\usepackage[pdfencoding=auto, psdextra]{hyperref}
\usepackage[normalem]{ulem}
\usepackage{txfonts}
\usepackage[normalem]{ ulem }
\usepackage{soul}
\usepackage{amsmath}
\usepackage{natbib}\usepackage{hyperref}
\hypersetup{
    colorlinks=true,
    linkcolor=blue,
    filecolor=magenta,      
    urlcolor=cyan,
    citecolor=blue
}

\newcommand{\msun}{{\rm M_\odot}}
\newcommand{\feh}{\rm{[Fe/H]}}
\newcommand{\teff}{T_{\rm eff}}
\newcommand{\logg}{\log{g}}
\newcommand{\vmic}{v_{\rm mic}}
\begin{document} 

\title{The Gaia-ESO survey: 3D NLTE abundances in the open cluster \object{NGC 2420} suggest atomic diffusion and turbulent mixing are at the origin of chemical abundance variations\thanks{Based on observations collected with the ESO Very Large Telescope at the La Silla Paranal Observatory in Chile for the Gaia-ESO Public Survey (program IDs 188.B-3002 and 193.B-0936).}}
%
\author{Ekaterina Semenova\inst{1,2}, Maria Bergemann\inst{1},  Morgan Deal\inst{3}, Aldo Serenelli\inst{4,5}, Camilla Juul Hansen\inst{1}, Andrew J. Gallagher\inst{1}, Amelia Bayo\inst{6}, Thomas Bensby\inst{7}, Angela Bragaglia\inst{8},  Giovanni Carraro\inst{9}, Lorenzo Morbidelli\inst{10}, Elena Pancino\inst{10}, Rodolfo Smiljanic\inst{11}}

\institute{ Max-Planck Institut f\"{u}r Astronomie, K\"{o}nigstuhl 17, 69117 Heidelberg, Germany\\
          \email{semenova@mpia.de}
\and 
Kazan Federal University, Kazan 420008, Russian Federation
\and
Instituto de Astrof\'{i}sica e Ci\^{e}ncias do Espa\c{c}o, Universidade do Porto CAUP, Rua das Estrelas, PT4150-762 Porto, Portugal\\
\email{morgan.deal@astro.up.pt}
\and
Institute of Space Sciences (ICE, CSIC), Carrer de Can Magrans S/N, E-08193, Cerdanyola del Valles, Spain
\and
Institut d'Estudis Espacials de Catalunya (IEEC), Carrer Gran Capita 2, E-08034, Barcelona, Spain
\and
Instituto  de  F\'{i}sica  y Astronom\'{i}a,  Facultad  de Ciencias,  Universidad de Valpara\'{i}so, Av. Gran Breta\~na 1111, 5030 Casilla, Valpara\'{i}so, Chile
\and
Lund Observatory, Department of Astronomy and Theoretical Physics, Box 43, SE-221 00 Lund, Sweden
\and
INAF - Osservatorio di Astrofisica e Scienza dello Spazio di Bologna, via Gobetti 93/3, 40129, Bologna, Italy
\and
Dipartimento di Fisica e Astronomia, Universit\`{a} di Padova, Vicolo dell'Osservatorio 3, 35122 Padova, Italy
\and
INAF - Osservatorio Astrofisico di Arcetri, Largo E. Fermi 5, 50125, Florence, Italy
\and
Nicolaus Copernicus Astronomical Center, Polish Academy of Sciences, 00-716, Warsaw, Poland
}

\date{Received 3 July 2020; accepted  2020}

  \abstract
   {Atomic diffusion and mixing processes in stellar interiors influence the structure and the surface composition of stars. Some of these processes cannot yet be modelled from the first principles, and they require calibrations. This limits their applicability in stellar models used for studies of stellar populations and Galactic evolution.}
   {Our main goal is to put constraints on the stellar structure and evolution models using new refined measurements of the chemical composition in stars of a Galactic open cluster.}
   {We used medium-resolution, $19\,200 \leq R \leq 21\,500$, optical spectra of stars in the open cluster NGC 2420 obtained within the Gaia-ESO survey. The sample covers all evolutionary stages from the main sequence to the red giant branch. Stellar parameters were derived using a combined Bayesian analysis of spectra, 2MASS photometry, and astrometric data from Gaia DR2. The abundances of Mg, Ca, Fe, and Li were determined from non-local thermodynamic equilibrium (NLTE) synthetic spectra, which were computed using one-dimensional (1D) and averaged three-dimensional (3D) model atmospheres. 
   We compare our results with a grid of Code d’Evolution Stellaire Adaptatif et Modulaire (CESTAM) stellar evolution models, which include atomic diffusion, turbulent, and rotational mixing.}
   {We find prominent evolutionary trends in the abundances of Fe, Ca, Mg, and Li with the mass of the stars in the cluster. Furthermore, Fe, Mg, and Ca show a depletion at the cluster turn-off, but the abundances gradually increase and flatten near the base of the red giant branch (RGB).
   The abundance trend for Li displays a signature of rotational mixing on the main sequence and abrupt depletion on the sub-giant branch, which is caused by advection of Li-poor material to the surface. The analysis of abundances combined with the CESTAM model predictions allows us to place limits on the parameter space of the models and to constrain the zone in the stellar interior, where turbulent mixing takes place.}
    {}

\titlerunning{Abundance variations in NGC~2420.}
\authorrunning{E. Semenova, et al.}

\maketitle
%

\section{Introduction}

Models of stellar evolution and their predictions in terms of nucleosynthesis in stars form the basis of many studies in modern astrophysics. Measurements of chemical abundances from stellar spectra provide the most detailed and accurate observational diagnostic of chemical composition of stellar atmospheres, and are, therefore,   routinely used in studies of chemical evolution of stellar populations in the Milky Way and other galaxies.

Until recently, it was common to assume that the abundances measured in the atmospheres of late-type (spectral type F, G, and K) stars reflected the composition of the material from which the stars were born. Selective modulations of surface abundances of Li, C, and N were known for red giant branch (RGB) stars and were canonically attributed to convective mixing and dredge-up episodes on the RGB \citep{Salaris2015}. Yet for most other evolutionary stages -- the main-sequence (MS), turn-off (TO), and sub-giant branch (SGB) -- strong evidence for distinct chemical signatures of self-processing in unevolved stars was lacking. This suggested that star clusters are simple mono-metallic stellar populations \citep[e.g.][]{Gratton2001, Thevenin2001, RamirezCohen2003}. These observations could not be reconciled with uncomfortably large effects of atomic diffusion -- a term that is nowadays used to refer to a combined action of gravitational settling and radiative acceleration -- theoretically predicted in early stellar structure calculations \citep[e.g.][]{Michaud1984, Turcotte1998, richard02}. As a consequence, it has become common to associate the measured abundance patterns with the variations in the chemical properties of the interstellar matter, disregarding the subtle yet important influence of secular effects in stellar evolution on the surface chemical composition of stars.

However, we are now witnessing a paradigm shift in the field, which is driven both by new observational studies and theoretical results. Empirical evidence of the impact of atomic diffusion on the surface chemical composition of stars is continuously emerging from careful observational studies of Galactic clusters with modern space- and ground-based  astronomical facilities \citep[e.g.][]{Korn2007, Gruyters2014, Blanco-Cuaresma2015, Gruyters2016, Husser2016, Gao2018, Motta2018, Souto2018, Liu2019, Souto2019}. For example, it is known that the abundances of light elements (Li, Be, and B) can be depleted in MS, TO, and SGB stars \citep{Smiljanic2010, Deliyannis2019, Boesgaard2020}, and these signatures have been linked to the effects of rotation-induced mixing, internal gravity waves, atomic diffusion, and thermohaline mixing. 
Large, statistically significant samples of stars with high-resolution spectra and high-quality astrometry (e.g. the Gaia-ESO survey: \citet{GES,Randich2013}; Gaia DR2: \citet{GaiaMission2016, GaiaDR22018}) probing the full evolutionary sequence from the lower MS to upper RGB are now available for many open clusters in the Milky Way. This allows unambiguous membership classification, accurate analysis of evolutionary stages of stars, and in turn, robust identification of systematic abundance variations along the evolutionary sequence of a cluster. One of the major results of the recent detailed  investigations is a systematic depletion, to the order of$\sim 0.15$ dex, of the abundances of light (Mg, Ca) and Fe-group elements at the TO of several Galactic open clusters when compared with their lower MS and RGB stars, which is qualitatively consistent with theoretical predictions \citep{Gao2018,Souto2018}.

Major progress with the implementation of non-standard chemical mixing processes in stellar structure models has been made over the past decade. These include micro- and macroscopic mixing processes. Microscopic mixing has a different impact on different chemical elements, and it includes gravitational settling, thermal diffusion, and radiative acceleration among other effects. In contrast, macroscopic processes, such as rotational and thermohaline mixing, act on all chemical species in the same way. Earlier theoretical studies of atomic diffusion and mixing in stellar structure calculations \citep{richard02, DeliyannisPinsonneault1990, Proffitt1991, Richer1992, Vauclair1999, Chaboyer2001, Richard2005} are now being superseded by the new generation of stellar evolution models \citep[e.g.][]{theado09,vick13,zhang19,deal20} that include both atomic diffusion and transport processes as thermohaline convection, mass loss, rotation, or accretion. A comprehensive review of the subject can be found in \citet{SalarisCassisi2017}.

Despite all the advances, major uncertainties in understanding of the physical mechanisms underlying the transport of elements in stellar interior remain. It has become clear that additional mixing processes, such as parametrised turbulent mixing, are necessary to reconcile observations with stellar evolution models \citep[e.g.][]{richer00,richard01,michaud11}. Also observational studies are still limited and provide only a fragmented picture of secular stellar evolution and its relation to abundances in stellar atmospheres. Most studies to date focus on the analysis of small stellar samples comprising no more than a dozen stars in each evolutionary stage, and they do not probe the critical regime of age and metallicity, where the combined effects of secular stellar evolution are expected to be at the maximum. M67 is the best studied system in this respect. However, this cluster is too old and its turn-off is too cool to reveal the subtle difference in the abundance patterns caused by diffusion processes in the interior \citep[e.g.][]{deal18}.

Motivated by the availability of new observational data, in this study we performed a detailed chemical abundance analysis of 70 stars in the open cluster \object{NGC 2420} (also known as Collinder 154, Melotte 69). This relatively young $\tau \sim 2$ Gyr \citep{Bossini2019} cluster was recently observed within the Gaia-ESO large spectroscopic stellar survey \citep{GES, Randich2013}. Accurate proper motions and parallaxes have also become available from the second data release (Gaia DR2) of the Gaia space mission \citep{GaiaMission2016, GaiaDR22018}. The observed sample of stars includes the full evolutionary sequence, from the lower MS to the RGB tip. \object{NGC 2420} is an ideal ensemble to study the effects of atomic diffusion, as it is relatively metal-poor (which maximises the effect of radiative acceleration), and it hosts early F-type stars with $T_{\rm eff} \approx 6500$ K at the TO region. These stars are luminous with very thin convective envelopes, and, as a consequence, the changes in surface composition caused by the combined effects of mixing, gravitational settling, and radiative acceleration, should remain easily detectable, in contrast to cooler G-type solar-like stars, which harbour more massive convective envelopes that efficiently mix the material and act as buffers to wash out the fine signatures of individual transport processes.

The paper is organised as follows. In Sect.~\ref{observations}, we present the observed data, while in Sect.~\ref{methods}, we describe methods used to compute stellar  parameters and provide a discussion of three-dimensional (3D) and non-local thermodynamic equilibrium (NLTE) effects. Section~\ref{models} summarises the key aspects of Code d’Evolution Stellaire Adaptatif et Modulaire (CESTAM, the 'T' stands for transport) stellar evolution models, which are employed to model the impact of atomic diffusion, turbulent mixing and rotation on the surface enrichment, and to interpret the observed depletion and accumulation of chemical elements. In Sect.~\ref{results}, we present the results of the abundance analysis and test them against the CESTAM models. We compare our findings with previous studies and discuss them in the context of stellar evolution and Galactic archaeology in Sect.~\ref{discussion}. 
In Sect.~\ref{perspectives}, we outline the future perspectives of our findings. Finally, Sect.~\ref{conclusions} provides a summary of our results.
\section{Observations}\label{observations}
The Gaia-ESO large spectroscopic survey was designed to obtain high-quality spectroscopic observations of $100\,000$ field stars, as well as members of clusters down to the limiting magnitude $\mathrm{V=19^{m}}$. For a further overview, the reader is referred to \citet{GES} and \citet{Randich2013}. In this work, we made use of spectroscopic data obtained with the medium-resolution  GIRAFFE spectrograph mounted at the Very Large Telescope (VLT). Although some targets in the cluster were also observed with the high-resolution UVES spectrograph, the spectra are only available for a few stars on the RGB. The pre-selection of targets in this cluster was based on the colour-magnitude diagram (CMD) from earlier photometric studies \citep{Anthony-Twarog2006, Sharma2006}. Post-processing of raw observed spectra was done by the Gaia-ESO dedicated work groups. We used the spectra released as part of the fifth internal Data Release (iDR5), which contains 545 objects labelled as cluster candidates. The  signal-to-noise ratio (SNR) of the spectra ranges from 10 to 150, and for our final sample the median SNR is 70.
\begin{figure}[h!]
\includegraphics[width=\linewidth]{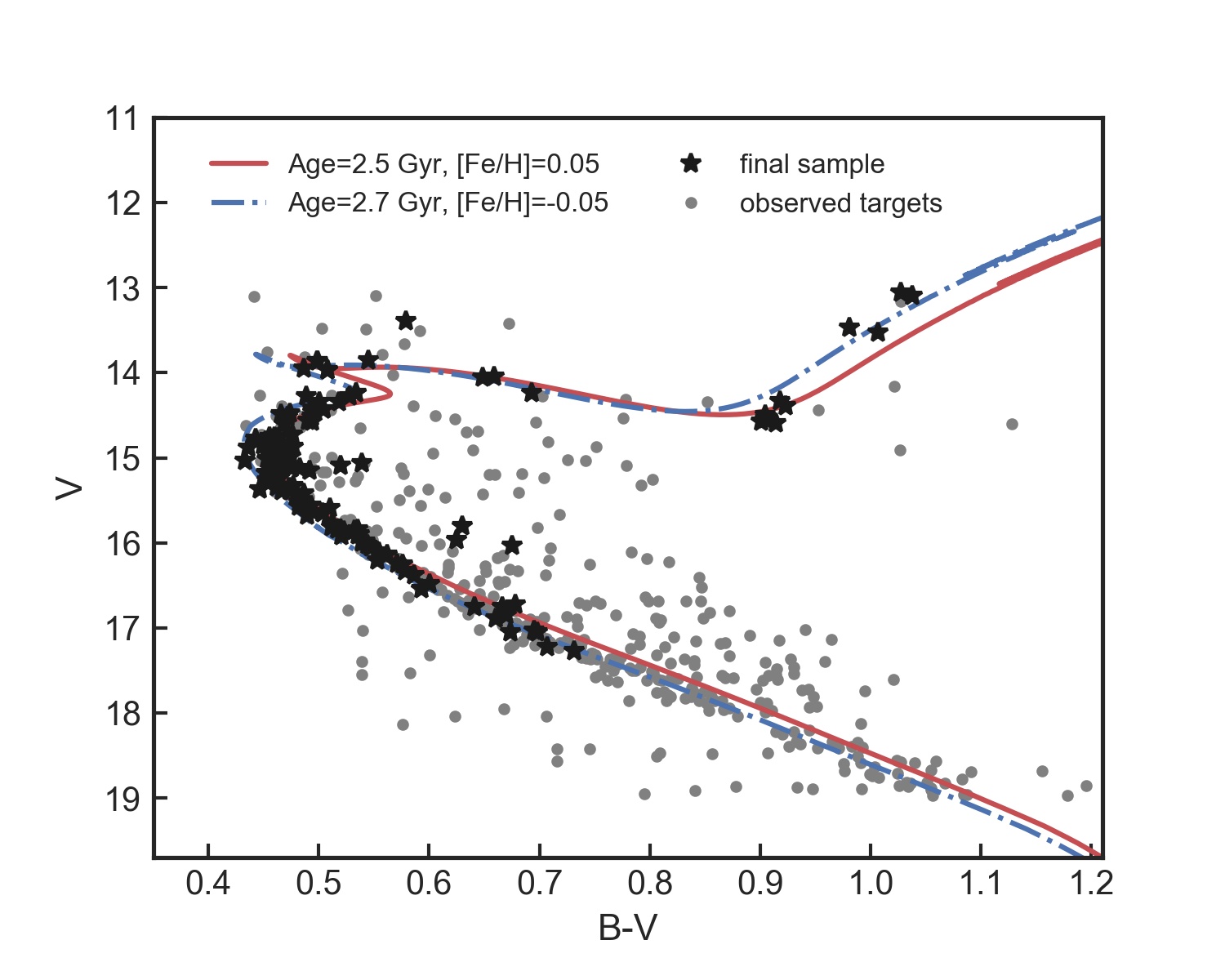}
\includegraphics[width=\linewidth]{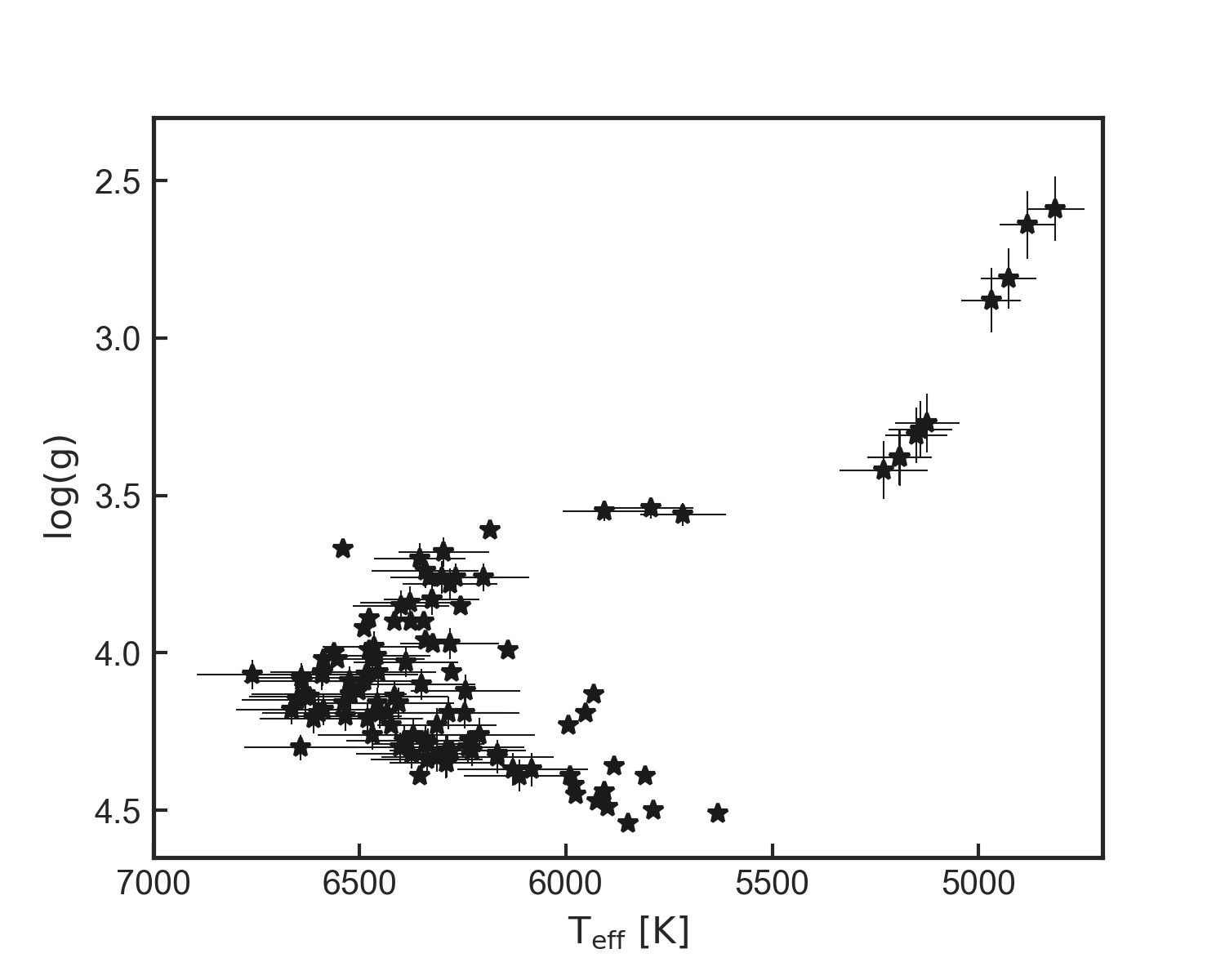}
\caption{Top: B-V colours of the observed stars in the open cluster NGC 2420 plotted as a function of their V magnitude. The best fit GARSTEC isochrones are over-plotted (see text). Stars comprising the final sample are depicted with black asterisks. For selection criteria, see main text. Some of the stars in the kinematically selected sample could be binaries. Bottom: Hertzsprung-Russell diagram of stars for which we performed a detailed spectroscopic analysis.}
\label{Fig:CMDfit}%
\end{figure}
\section{Methods}\label{methods}
\subsection{Target selection}
The recent Gaia DR2 estimate of the cluster parallax, $\pi = 0.363 \pm 0.064$ milli-arcsec (mas), yields the distance of $2.55$ kpc with an uncertainty of about $0.5$ kpc \citep{Cantat-Gaudin2018}. Owing to the large distance, we refrained from carrying out membership analysis based on the proper motions of the stars and instead selected cluster members by their apparent positions and radial velocities. We required the radial velocity to be within the range of $73-77\, \mathrm{km/s}$ and angular distance from the cluster centre to be under 10 arcmin, according to the observable size of the cluster as reported in \citet{Sharma2006}. This procedure effectively eliminated 214 foreground and background stars and yielded 331 cluster candidate members. We note that adding the proper motions in the analysis does not change our classification, as the uncertainties of proper motions are large. We explicitly avoided pre-selection by metallicity, as stellar structure models computed with atomic diffusion and mixing  predict a dispersion in the chemical composition of the cluster, hence any pre-selection based on the chemical composition significantly threatens to erase these astrophysically important signatures that are the focus of our work. 

Gaia photometry of the cluster reveals a characteristic broadening of the cluster MS. One of the viable explanations for this feature is the presence of unresolved binaries in our sample. it is a well-established fact  \citep[e.g.][]{Maeder1974, Bragaglia2006, Cordoni2018, ElBadry2018, Price-Whelan2020} that the unresolved binaries composed of two MS stars are redder and brighter than single MS stars of a similar mass. These unresolved binary systems may appear up to $\sim$ 0.75 mag brighter than the canonical main sequence that characterises evolution of single stars. However, it is not only the visual brightness, but also the colour as a proxy for $\teff$ that is affected \citep{ElBadry2018}. The effects become significant for binaries in which both components have similar masses \citep{ElBadry2018}. According to the statistical method by \citet{Cordoni2018}, NGC~2420 has a significant (33~\%) fraction of unresolved MS-MS binaries. Binaries with mass ratios $q>0.7$ constitute 10 \% of NGC~2420 members. We therefore excluded those stars that have a high likelihood (based on the CMD position) of being binaries from the subsequent analysis.
\subsection{Age of the cluster \label{age}}
We estimated the age of the cluster by fitting the observed Johnson-Cousins photometry \citep{Sharma2006} and Gaia parallaxes\footnote{We accessed the parallaxes available in the Gaia DR2 using the ASTROPY community Python library.} to a grid of stellar isochrones, as described below. The cluster is almost unaffected by reddening. According to the NASA/IPAC Infrared Science Archive service,\footnote{\url{irsa.ipac.caltech.edu/applications/DUST/}} E(B-V) $= 0.035$ mag, in agreement with earlier studies \citep{Anthony-Twarog2006}. NGC~2420 has been considered to be a moderately metal-deficient open cluster, being a 'transition' object between the solar metallicity open clusters and more metal-poor globular clusters. Some of the recent studies targeting members of NGC~2420 report the average metallicity of the cluster to be at $ \feh = -0.05 \pm 0.10$ \citep[][based on a few stars observed at high resolution]{Pancino2010} or $ \feh = -0.2 \pm 0.06$ \citep[][based on stars observed with a medium-resolution spectrograph]{Jacobson2011}. \cite{Siegel2019} suggested that isochrones with lower than previously found metallicity (by $-0.1$ dex) are needed to describe the photometry of the cluster TO stars.

We used the grid of GARSTEC stellar isochrones \citep{Weiss2008}, which are based on the same stellar models as used in BeSPP \citep{serenelli:2017}. Synthetic photometry was computed using bolometric corrections based on ATLAS12/SYNTHE \citep{Kurucz1970,Kurucz1993} as implemented by Conroy et al. (in prep.).\footnote{\url{http://waps.cfa.harvard.edu/MIST/model_grids.html\#bolometric}} Zero point corrections were applied to reproduce the solar colours from \citet{Casagrande2018}. 

To break the age-metallicity degeneracy, we assumed the metallicity of the cluster to be the one of the most evolved stars in our sample, which are located on the RGB. Metallicity of those stars was computed with spectrum synthesis (see Sect. \ref{fundamentalparameters} for more details).
Figure \ref{Fig:CMDfit} shows two GARSTEC isochrones, which correspond to the age of $2.5$ and $2.7$ Gyr, respectively. We caution that the standard procedure of fitting the grid of isochrones to the CMD, although widespread in astronomy (see i.e. \citet{PontEyer2004}), assumes that the cluster is a mono-metallic coeval system. In fact, this contradicts our findings (Sect. \ref{results}) of a systematic depletion of elemental abundances at the cluster TO point, which we interpret as a signature of atomic diffusion and mixing. However, employing a different, stricter approach is not feasible at this stage. Indeed, a systematic depletion of metallicity in principle requires an iterative procedure involving the full analysis of the observed spectra and Bayesian stellar evolution fitting. Developing such a model is beyond the scope of our study. However, in the next section we show that fundamental stellar parameters are not affected at any significant level by the assumed average metallicity of the cluster.
\subsection{Stellar parameters and chemical abundances}\label{fundamentalparameters}
We used several methods to constrain stellar parameters: analysis of photometry and parallaxes, fitting of the Balmer lines, and the full Bayesian approach employing stellar evolution models and parallaxes. All these methods are broadly used in the literature and have been verified in our previous studies with different families of synthetic spectral models \citep{Bergemann2012, Ruchti2013, Serenelli2013}. We followed this approach rather than using the recommended Gaia-ESO parameters and abundances because it allowed us to remain fully consistent with our one-dimensional (1D) NLTE and 3D NLTE calculations. First, 3D NLTE abundances are the quantities that we use for astrophysical interpretation (Sect. 5). Second, this allows an objective analysis of systematic and statistical uncertainties, which are associated with every step in stellar parameter determinations. 

Photometric $\teff$ were derived from the $(V-K)$ colour using the \citet{Alonso1996} and \citet{Casagrande2010} calibration relations. We assumed the same metallicity, $\feh= -0.2$, for all cluster members, but we verified that the variation of metallicity had no significant impact on the $\teff$ estimates. These estimates of $\teff$ were then employed as initial guesses for the spectroscopic analysis of $H_{\alpha}$ line wings (see \citet{Ruchti2013} for the details of this method). 

\begin{figure}[h!]
\centering
\includegraphics[width=\linewidth]{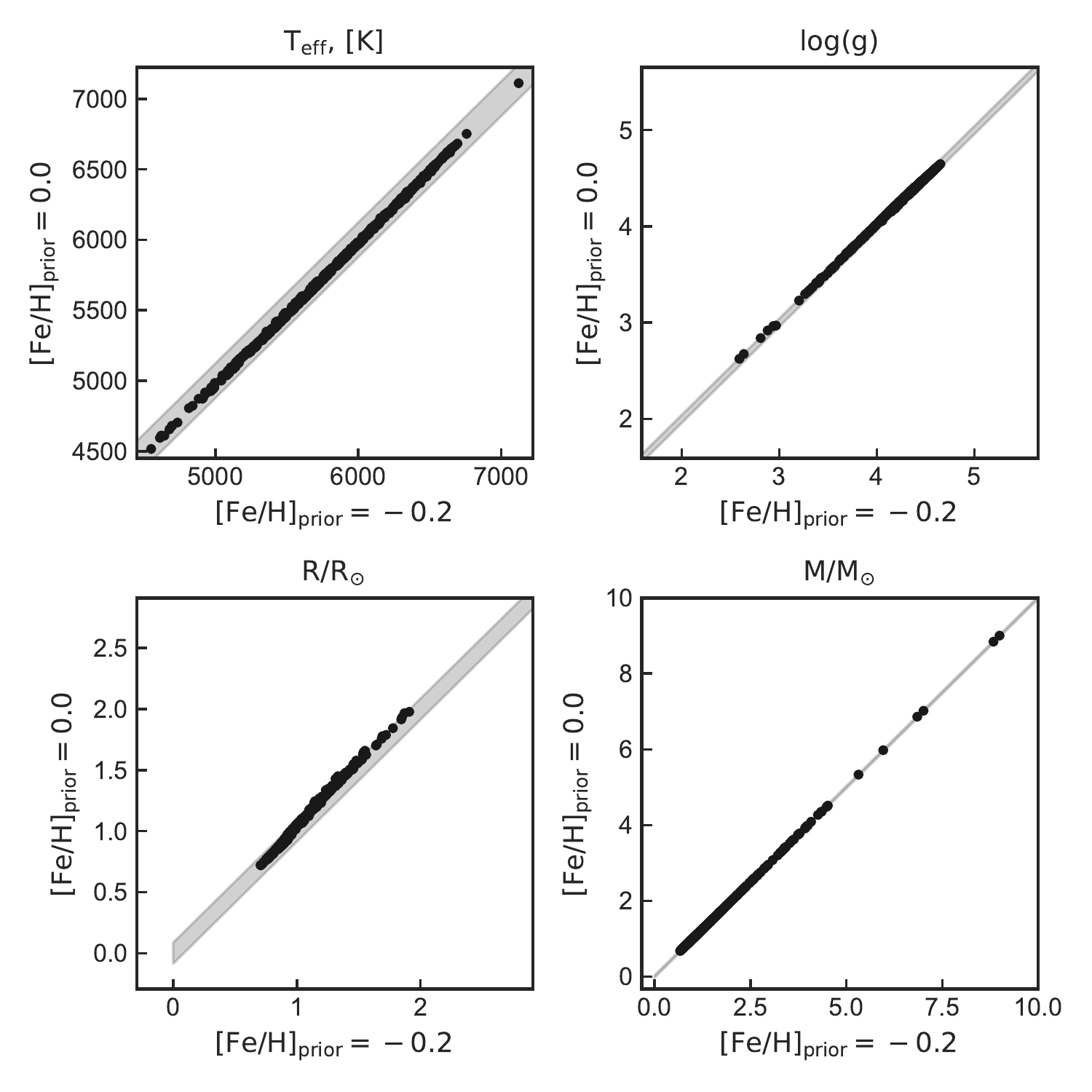}
\caption{Comparison of $\teff$, $\log{\mathrm{(g)}}$, masses, and radii derived from the Bayesian analysis as described in \citet{Serenelli2013} assuming a uniform metallicity of a cluster [Fe/H] = 0 or [Fe/H] = -0.2. Shaded ares depict the range of mean errors on derived parameters, $\sigma (\mathrm{T_{eff}}) = 120\, \mathrm{K}$, $\sigma (\log{g}) = 0.05\, \mathrm{dex}$, $\sigma (\mathrm{R/ R_{\odot}}) = 0.1$, $\sigma (\mathrm{M / M_{\odot}}) = 0.05,$ respectively. Thus we show that the derived fundamental parameters are not biased by the initial guess of the metallicity.}
\label{parameters_vs_FeH}%
\end{figure}

We used the SME 1D code under the assumption of local thermodynamic equilibrium (LTE) \citep{SME} and MARCS model atmospheres \citep{MARCS} to generate synthetic model spectra and fit them to the observed spectra. Finally, we resorted to the Bayesian code BeSPP \citep{Serenelli2013} to refine our photometric and spectroscopic estimates of $\teff$, and to derive estimates of $\log g$ for our targets. Assuming the Gaussian uncertainty of $\pm150~$K on the spectroscopic values, we combined them with the 2MASS $JHK$ magnitudes, Gaia parallaxes, and we adopted a uniform metallicity prior ($\feh=-0.20\pm0.30$). The final estimates of $\teff$ and $\log g$ were determined from the analysis of the full posterior probability distribution functions (PDFs) as described in \citet{Serenelli2013}. We note that the final estimates are not affected in any significant way by assuming a uniform metallicity for the cluster (see Fig. \ref{parameters_vs_FeH}).

The analysis of metallicities and chemical abundances for individual stars is strictly spectroscopic, and we rely on the method of detailed spectrum synthesis. Although the GIRAFFE HR10 and HR15N spectra cover only a limited wavelength range, we have relatively unblended spectrum
features of $15$ Fe I and $2$ Fe II, as well as a few clean features of other chemical elements that are suitable for a high-quality abundance analysis. The parameters of these lines are provided in Table \ref{table:spectralLines}.

\begin{table}
\caption{Main properties of spectral features used to compute stellar abundances with SME.}
\label{table:spectralLines}
\centering          
\begin{tabular}{c c c c c c c}   
\hline\hline    
Element & Ion  & $\lambda, \AA$ & $\log{(\rm gf)}$& $\mathrm{E_{low}}$, eV & $\mathrm{E_{up}, eV}$ \\
\hline 
Fe & $ 1 $ &  $ 5339.929 $  &  $  -0.667   $ & $       3.266  $ & $ 5.587  $ \\
Fe & $ 1 $ &  $ 5364.871 $  &  $   0.228   $ & $       4.446  $ & $ 6.756  $ \\
Fe & $ 1 $ &  $ 5373.709 $  &  $  -0.760   $ & $       4.473  $ & $ 6.780  $ \\
Fe & $ 1 $ &  $ 5379.574 $  &  $  -1.514   $ & $       3.695  $ & $ 5.999  $ \\
Fe & $ 1 $ &  $ 5389.479 $  &  $  -0.410   $ & $       4.415  $ & $ 6.715  $ \\
Fe & $ 1 $ &  $ 5393.167 $  &  $  -0.715   $ & $       3.241  $ & $ 5.539  $ \\
Fe & $ 1 $ &  $ 5398.279 $  &  $  -0.630   $ & $       4.446  $ & $ 6.742  $ \\
Fe & $ 1 $ &  $ 5434.524 $  &  $  -2.121   $ & $       1.011  $ & $ 3.292  $ \\
Fe & $ 1 $ &  $ 5445.042 $  &  $  -0.020   $ & $       4.387  $ & $ 6.663  $ \\
Fe & $ 1 $ &  $ 5506.779 $  &  $  -2.795   $ & $       0.990  $ & $ 3.241  $ \\
Fe & $ 1 $ &  $ 5560.212 $  &  $  -1.090   $ & $       4.435  $ & $ 6.664  $ \\
Fe & $ 1 $ &  $ 5587.574 $  &  $  -1.750   $ & $       4.143  $ & $ 6.361  $ \\
Fe & $ 1 $ &  $ 6494.980 $  &  $  -1.268   $ & $       2.404  $ & $ 4.313  $ \\
Fe & $ 1 $ &  $ 6593.870 $  &  $  -2.420   $ & $       2.433  $ & $ 4.313  $ \\
Fe & $ 1 $ &  $ 6710.318 $  &  $  -4.764   $ & $       1.485  $ & $ 3.332  $ \\
Fe & $ 2 $ &  $ 5425.249 $  &  $  -3.220   $ & $       3.199  $ & $ 5.484  $ \\
Fe & $ 2 $ &  $ 6516.077 $  &  $  -3.310   $ & $       2.891  $ & $ 4.793  $ \\
~\\ 
Ca & $ 1 $ &  $ 5349.465 $  &  $  -0.310   $ & $       2.709  $ & $ 5.026  $ \\
Ca & $ 1 $ &  $ 5512.980 $  &  $  -0.464   $ & $       2.933  $ & $ 5.181  $ \\
Ca & $ 1 $ &  $ 5581.965 $  &  $  -0.555   $ & $       2.523  $ & $ 4.744  $ \\
Ca & $ 1 $ &  $ 5588.749 $  &  $   0.358   $ & $       2.526  $ & $ 4.744  $ \\
Ca & $ 1 $ &  $ 5590.114 $  &  $  -0.571   $ & $       2.521  $ & $ 4.739  $ \\
Ca & $ 1 $ &  $ 5594.462 $  &  $   0.097   $ & $       2.523  $ & $ 4.739  $ \\
Ca & $ 1 $ &  $ 5601.277 $  &  $  -0.523   $ & $       2.526  $ & $ 4.739  $ \\
Ca & $ 1 $ &  $ 5602.842 $  &  $  -0.564   $ & $       2.523  $ & $ 0.000  $ \\
Ca & $ 1 $ &  $ 6471.662 $  &  $  -0.686   $ & $       2.526  $ & $ 4.441  $ \\
Ca & $ 1 $ &  $ 6499.650 $  &  $  -0.818   $ & $       2.523  $ & $ 4.430  $ \\
~\\
Mg & $ 1 $ &  $ 5528.405 $  &  $  -0.498   $ & $       4.346  $ & $ 6.588  $ \\
~\\
Li & $ 1 $ &  $ 6707.764 $  &  $  -0.002   $ & $       0.000  $ & $ 1.848  $ \\
Li & $ 1 $ &  $ 6707.915 $  &  $  -0.303   $ & $       0.000  $ & $ 1.848  $ \\
Li & $ 1 $ &  $ 6707.922 $  &  $  -1.122   $ & $       0.000  $ & $ 0.000  $ \\
Li & $ 1 $ &  $ 6708.073 $  &  $  -1.423   $ & $       0.000  $ & $ 0.000  $ \\
\hline                  
\end{tabular}
\end{table}
All atomic data are adopted from the official Gaia-ESO line list (see \citealt{Heiter2015} for details). We note that some species (Mg and Li) are represented by one spectral feature in our observed spectral data. We have, therefore, taken special care to assess all sources of error in the abundance analysis of the diagnostic features, including statistical and systematic uncertainties. The assumption of 1D LTE is arguably the most severe source of systematic error in abundance estimates \citep[e.g.][]{Asplund2005, BergemannNordlander2014}. Furthermore, 3D and NLTE effects are functions of the evolutionary stage. We therefore performed detailed calculations of NLTE abundances using canonical 1D hydrostatic model atmospheres and 3D hydrodynamic model atmospheres. The detailed approach to NLTE computations is described in the following sections.
\subsubsection{1D NLTE abundances}\label{1DNLTE}
The 1D statistical equilibrium code MULTI2.3 \citep{Carlsson1986} is used to compute grids of 1D LTE and NLTE line profiles, and, consequently, NLTE abundance corrections, $\Delta_{\rm NLTE}$, via interpolation in the LTE and NLTE curves of growth as described in \citep{Eitner2019}. NLTE abundance correction describes the difference in abundance required to match a spectral line of a fixed equivalent width with LTE and NLTE models. This quantity generally depends on the atomic properties of the line, elemental abundance, and physical parameters of a stellar atmosphere. If $\Delta_{\rm NLTE}$ is known, the NLTE abundance of an element $A(X)_{\rm NLTE}$ is computed according to Eq.~\ref{eq:NLTEabundance}:
\begin{equation}
\label{eq:NLTEabundance}
A(X)_{\rm NLTE}  =  A(X)_{\rm LTE} + \Delta_{\rm NLTE}.\end{equation}
The NLTE correction $\Delta_{\rm NLTE}$ is positive when the NLTE line profile is weaker than its LTE equivalent. Vice versa, $\Delta_{\rm NLTE} < 0$ implies that the NLTE line profile is stronger than its LTE counterpart, given all other parameters in calculations (abundance, model atmosphere parameters) are identical. In the latter case, the LTE abundance is higher compared to the NLTE abundance. It should be stressed, however, that Fe is the only element for which $\Delta_{\rm NLTE}$ is strictly differential: the input parameters of the model atmosphere in LTE and NLTE calculations are the same. The NLTE abundance corrections for the other chemical elements - Mg, Ca, and Li - were computed using NLTE-corrected metallicities, which implies, for a given $\teff$,  $\logg$, and $\vmic$, that we correctly took into account the  second-order dependence of their NLTE correction on that of Fe. The NLTE corrections on Fe lines typically amount to $+0.03$ to $+0.07$ dex, which implies the sensitivity of the NLTE corrections for Mg and Ca is to the order of $0.01$ dex.

Background opacity tables for each of these elements were computed using the Turbospectrum radiative transfer code \citep{TurboSpectrum}. In 1D calculations, we used MARCS model atmospheres.
Statistical equilibrium  computations are performed under the trace element assumption for $135$ model atmospheres within a broad range of stellar parameters: $\teff=[4500, 7000]\,\mathrm{K},\,\logg=[2.5, 4.5],\,\feh=[-0.75, 0.25]$. The microturbulence parameter ranges within $1-1.2\, \mathrm{km/s}$.
\begin{figure}[h!]
\centering
\includegraphics[width=\linewidth]{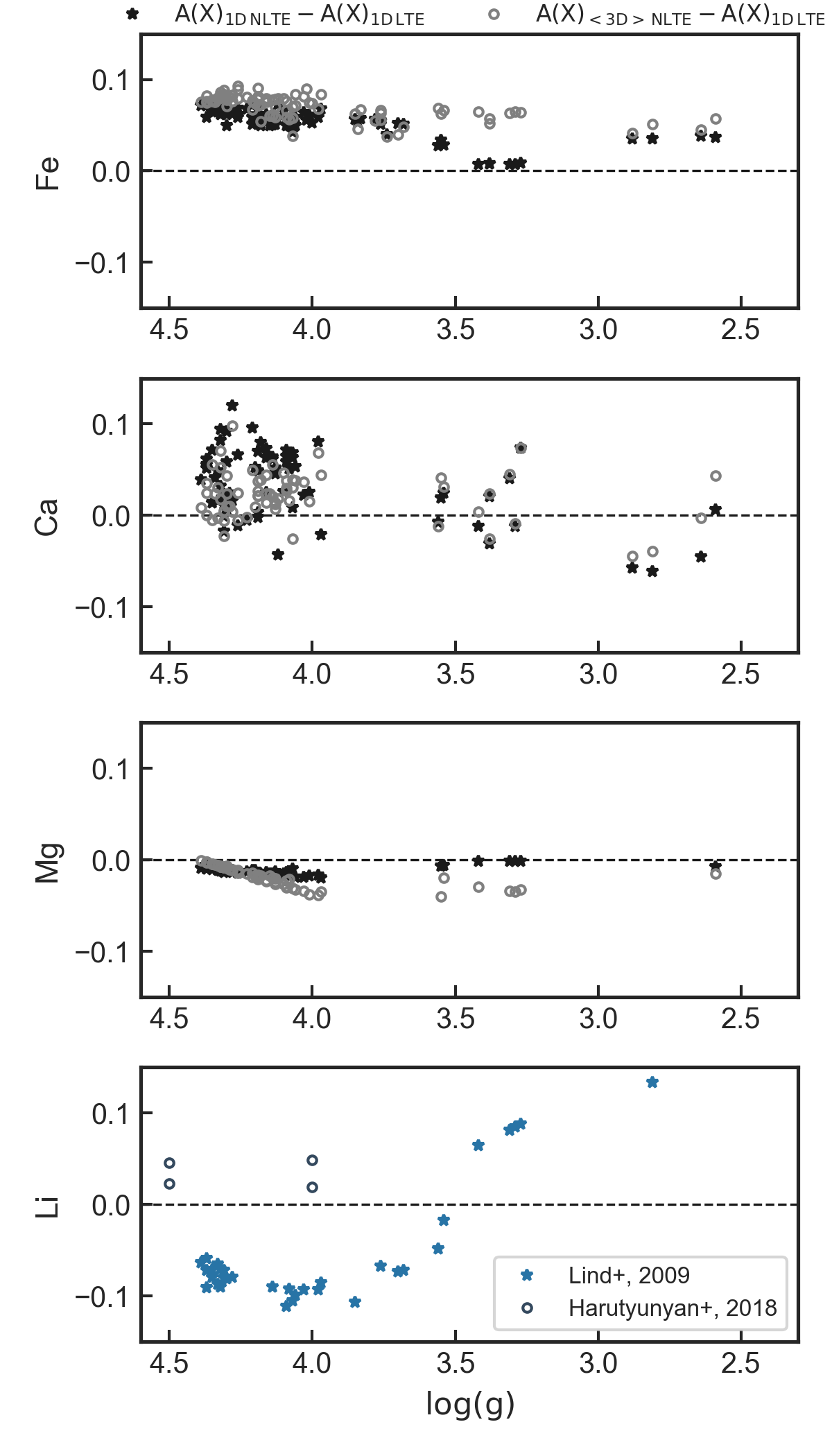}
\caption{Difference in abundances of investigated elements if derived in 1D NLTE and 3D NLTE vs. 1D LTE approach for the final sample of stars. Depicted difference is defined as $\Delta_{M} = A(X)_{M} - A(X)_{1D\, LTE}$, where M is the 3D NLTE or 1D NLTE case.}
\label{NLTEmLTE}
\end{figure} 
The atomic models for Mg and Ca are based on the models presented in the earlier studies by \citet{BergemannMg2017, Bergemann2012} and \citet{Mashonkina2017}. Table \ref{table:NLTEatoms} summarises the main properties of the atomic models, such as the number of energy levels and ionisation stages, the number of bound-bound, and the size of the frequency grid for radiative bound-free transitions. In this work, we update the reaction rates and cross-sections to more recent estimates available in the literature. 
In particular, we included the new photo-ionisation cross-sections for Fe I from \citet{Bautista2017} and replaced the semi-classical recipes for the rates of bound-bound and bound-free transitions cased by inelastic collisions between Fe$+$H \citep{BarklemFeH} and Fe$+$e \citep{Bautista2017}. In the Ca model atom, we updated rates for transitions caused by inelastic Ca$+$H collisions \citep{Belyaev2017CaH} and updated the list of energy levels to include fine structure resolved levels, and we therefore updated radiative bound-bound transitions. The estimates of line broadening caused by elastic collisions with H atoms are taken from \citet{BarklemPiskunovOMara}. We also reduced the complexity of the atomic models, in order to use them in the 3D NLTE calculations (Sect. \ref{3DNLTE}). We cut the Mg photo-ionisation cross-sections at 1100 $\AA,$\, as radiative fluxes at bluer wavelengths are negligibly small. For Fe, we re-sampled the photo-ionisation cross-sections, so that they consisted of a factor of $\sim 10$ fewer frequency points, but still contained all the important resonances. Numerous tests have been carried out to ensure that the atomic models with reduced complexity do not introduce any biases in 1D and 3D NLTE abundance corrections with respect to the original models. For Li, we made use of 1D NLTE corrections published by \citet{LindLi2009}. 
\begin{table*}
\caption{Main properties of atomic models used in statistical equilibrium and NLTE abundance calculations.}
\label{table:NLTEatoms}
\centering          
\begin{tabular}{l|ccc|cc|cc}   
\hline\hline      
Element & \multicolumn{3}{c}{\# Energy levels} & \multicolumn{2}{c}{\# Radiative transitions} & \multicolumn{2}{c}{\# Frequency points} \\
\hline 
   Fe & 548 Fe I & 58 Fe II & 1 Fe III & 12 635 b-b & 606 b-f & 89 537 b-b& 64 455 b-f \\ 
  Ca & 67  Ca I & 37 Ca II & 1 Ca III & 624 b-b & 104 b-f & 7 520 b-b& 10 433 b-f \\
  Mg & 85  Mg I & 1 Mg II &X Mg III & 246 b-b & 85 b-f & 2 940 b-b& 20 242 b-f \\
\hline                  
\end{tabular}
\end{table*}

Our 1D NLTE abundance corrections are shown in Fig. \ref{NLTEmLTE}. The NLTE corrections for the optical Fe I lines are moderate and do not exceed 0.15 dex, supporting previous estimates in the literature \citep{Bergemann2012, Ezzeddine2018}. For the Mg I 5528 \AA\ line, the NLTE correction is small: it varies from 0.03 dex for the TO model to $< 0.01$ for the main-sequence and sub-giant models. Ca I lines are typically weaker in NLTE compared to LTE, therefore, the NLTE corrections are moderately positive and range from 0.05 dex on the main sequence to $-0.1$ dex on the sub-giant branch. NLTE corrections to the Li I 6707 line are negative for main-sequence models, but become positive on the RGB. We emphasise that in virtue of the metallicity dependence of the NLTE abundance corrections for every element other than Fe, our estimates of NLTE effects for the NGC~2420 stars may not be directly comparable with other studies.
\subsubsection{3D NLTE abundances}\label{3DNLTE}
As a proxy of 3D structure of stellar atmospheres, we used averaged 3D atmospheric models taken from the STAGGER grid of stellar convection simulations \citep{Magic2013b}. The procedure to compute the 3D NLTE corrections is identical to that used for 1D NLTE calculations (\ref{1DNLTE}) and it follows our approach in \citet{Bergemann2012}. To account for 3D convective motions, we include turbulent velocity in otherwise 1D model atmosphere according to Eq. \ref{vturbMean3D} as proposed by \citet{UitenbroekCriscuoli2011}:
\begin{equation}
V_{\rm turb} = \frac{1}{3}\sqrt{<v^2> -(<v_x>^2 + <v_y>^2 + <v_z>^2)}
\label{vturbMean3D}
.\end{equation}
We emphasise that the 3D models do not include any parametrisations of convection, and so there are no ad-hoc  parameters, such as the mixing length or micro- and macro-turbulence, which allowed us to perform radiation transfer from the first principles.

The estimates of $\Delta_{\rm 3D NLTE}$ for Fe, Ca, and Mg are shown in Fig.\,\ref{NLTEmLTE}. Our results for 3D NLTE corrections compare favourably with the earlier estimates in the literature. For Ca, this is the first study of 3D NLTE effects for stars other than the Sun.
For Fe and Mg, the 3D NLTE corrections are within $0.05$ dex, which is consistent with the calculations by \citet{Bergemann2012} and \citet{BergemannMg2017}, respectively. 

Our results for Ca in 1D NLTE are consistent with \citep{Mashonkina2017}, who predicted modest and positive NLTE abundance corrections. The 3D effects, however, are insignificant. For Li we estimate 3D NLTE corrections based on the pre-computed grid by \citep{Harutyunyan2018}. At $\feh=-0.5,$ they predicted the 3D NLTE correction to be between 0.019 and 0.049 for MS and TO stars. We note that this value is much smaller than our observational errors ($\sim 0.1$ dex) and 1D NLTE effects ($\sim -0.15$ to $0.15$ dex).

The 3D NLTE corrections were applied to abundances measured in 1D LTE to derive the final chemical abundance pattern of \object{NGC 2420}. A more detailed investigation of the NLTE effects for the studied elements and a technical description of computations will be presented in the follow-up paper (Semenova et. al, in prep.).
\subsubsection{Propagation of uncertainties}\label{uncertainties}
For all measured chemical elements, we propagated corresponding abundance uncertainties using a Monte Carlo approach. We constructed a set of randomly sampled input parameters with respect to their errors and performed the spectrum fitting procedure with these input parameters fixed. The resulting error on derived abundances is defined as the standard deviation of the set of solution assuming a Gaussian distribution. 
Therefore, the errors presented account for fluctuations of the solutions due to the noise component (SNR of observed spectra) and systematic component (i.e. normalisation procedure and errors of fundamental parameters).
\section{Stellar evolution models with atomic diffusion and mixing}
\label{models}
\subsection{Stellar structure code}
The primary goal of this paper is to study the radial transport of chemical elements in stellar interiors. This physical phenomenon is caused by the competition and coupling between atomic diffusion and macroscopic transport processes. Atomic diffusion represents a balance between gravitational settling and radiative levitation forces \citep{Meynet2004}. The underlying mechanism is well understood, and the accuracy with which it can be modelled in stellar structure is roughly 20\%: this being limited by approximations in the atomic diffusion formalism and
 cross-sections see \citealt{michaud15} for a detailed overview. Other macroscopic transport processes of diffusive (e.g. rotation, thermohaline convection) or advective (e.g. mass loss) nature, are not yet well constrained and, therefore, their effects are usually modelled using simple parametric recipes \citep[e.g.][]{richer00,richard01,michaud11}. These processes can mitigate the effects of atomic diffusion on surface abundances.

We used the CESTAM code \citep{morel08,marques13,deal18} to compute stellar evolutionary tracks including atomic diffusion and additional parametrised mixing to provide quantitative predictions for the behaviour of the surface chemical composition for different evolutionary phases of a star. The code adopts the OPAL2005 equation of state \citep{rogers02} and the OP opacity tables \citep{seaton05}. These are complemented by the Wichita opacity data at low temperatures \citep{ferguson05}. We used opacity tables for a fixed solar mixture. We verified that the error related to this assumption (not recomputing the Rosseland mean opacity taking into account mixture variations due to atomic diffusion) was never larger than $0.01$~dex. This is due to the variations of chemical element abundances being relatively small, as they are inhibited by an additional transport process in our models.

Nuclear reaction rates are adopted from the NACRE compilation \citep{angulo99}, except for the $^{14}\mathrm{N}(p,\gamma)^{15}\mathrm{O}$ reaction, which is taken from the Laboratory for Underground Nuclear Astrophysics (LUNA) compilation \citep{imbriani04}. We used the Canuto-Goldman-Mazzitelli formalism for convection \citep{canuto96}, with the mixing-length parameter $\alpha_{CGM}=0.68$ calibrated on the Sun. The models also include the core overshoot of 0.15 $H_{\rm P}$. The stellar atmosphere, which represents the outer boundary condition in stellar structure models, is computed in the grey approximation. Following the recommendation of \citet{serenelli10}, we adopted the AGSS09 \citep{asplund09} mixture for the refractory elements. The initial hydrogen, helium, and metal mass fractions - $X_0$, $Y_0,$ and $Z_0,$ respectively - were determined from the solar calibration, following the $\Delta {\rm Y}/\Delta {\rm Z}$ slope of $0.9$ as determined in \citet{deal18}. We did not take mass loss or magnetic fields into account. The evolution is computed from the pre-main-sequence up to the age of 2.5~Gyr.
\subsection{Modelling transport processes}\label{sec:42}
\begin{figure*}[ht]
\centering
\includegraphics[width=\linewidth]{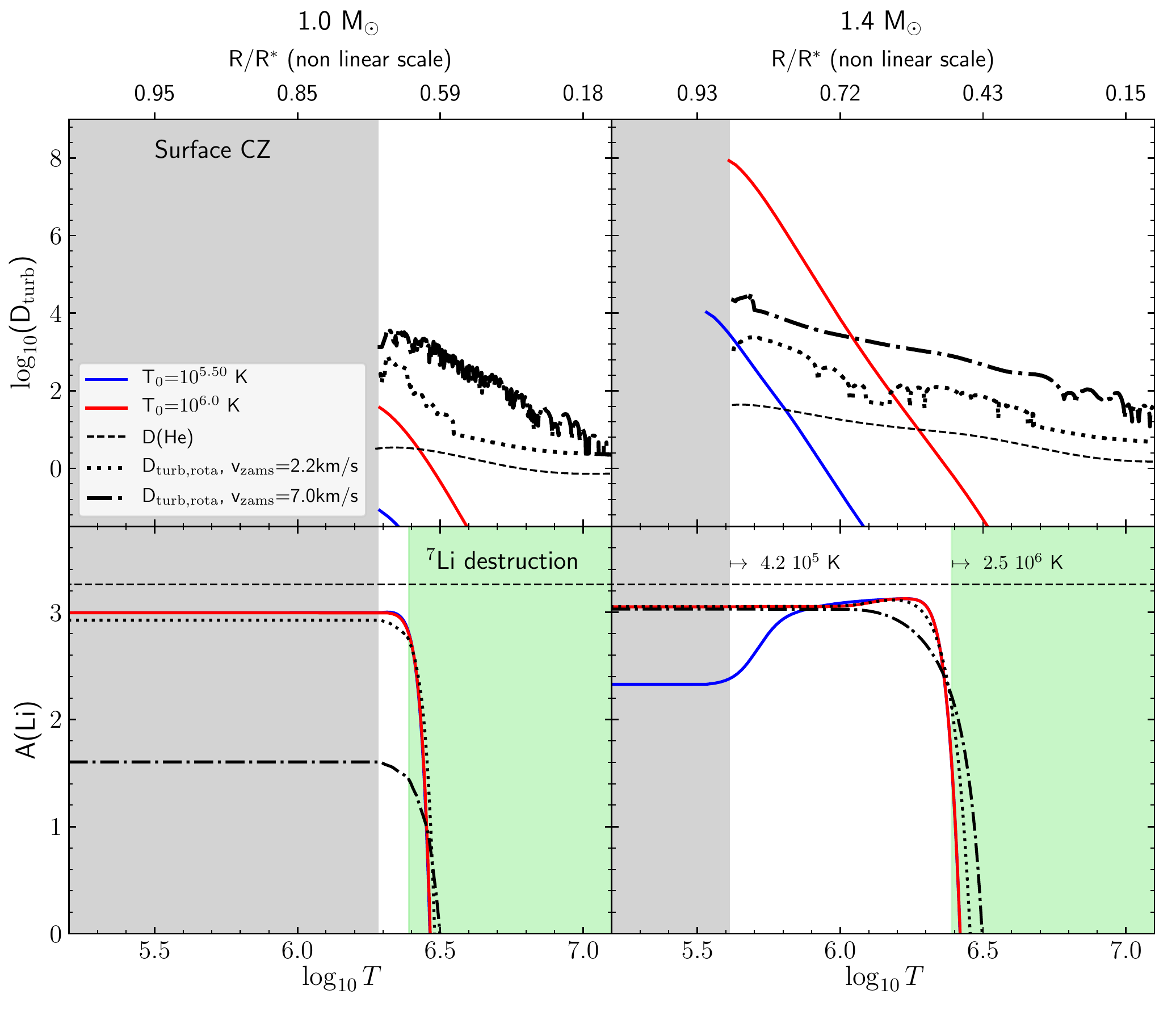}
\caption{Depth-dependent profiles of the atomic diffusion coefficient according to Eq. \ref{eqturb}  (upper panels) and Li abundance (lower panels) as a function of local kinetic temperature for two stellar models with masses of 1.0 $\msun$ (left panels) and 1.4 $\msun$ (right panel). Both models have an initial metallicity of [Fe/H] $=-0.05$ dex and an age of 2.5 Gyr. The profile of the atomic diffusion coefficient of He, commonly used as a reference value, is depicted with a dashed line. The blue and red lines correspond to different choices of the reference temperature T$_{0}$ (see also Eq. 5) resulting in different turbulent diffusion coefficients, and therefore in more (red line) or less (blue line) efficient turbulent mixing. The profiles of rotational mixing, computed using the zero-age main-sequence velocity of $\upsilon = 2.2$ and $7.0$ km/s, are indicated with the dotted and dash-dotted lines. On the lower panels, the same profiles are shown in the units of predicted Li abundance, depicting how Li abundance would change if affected by the prescribed mixing.
The grey shaded areas represent the sub-surface convective zone and the green shaded areas mark the inner regions (at $R/R* \lessapprox 0.6$ (1 $\msun$) and $R/R* \lessapprox 0.5$ (1.4 $\msun$)) where Li is destroyed by nuclear reactions.
For more details, we refer the reader to Sect.\,\ref{sec:s43} and Sect.\,\ref{sec:42}.}
\label{dDturbplot}
\end{figure*}
The diffusion equation of a trace element $i$ at a given depth is expressed as follows:
\begin{equation}\label{equadiff}
\rho\frac{\partial X_i}{\partial t}=\frac{1}{r^2}\frac{\partial}{\partial r}\left[r^2\rho {D_\mathrm{turb}}\frac{\partial X_i}{\partial r}\right]-\frac{1}{r^2}\frac{\partial}{\partial r}[r^2\rho v_{i}] + A_i m_p \left[\sum_j (r_{ji} -  r_{ij}) \right],
\end{equation}
where $X_i$ is the mass fraction of element $i$; $A_i$ its atomic mass; $m_p$  the mass of a proton; $\rho$ the density in the considered layer; $D_{\rm turb}$ the turbulent diffusion coefficient; and $r_{ij}$ the rate of the reaction that transforms the element $i$ into $j$. The competition between macroscopic transport processes and atomic diffusion is given by the first two terms in the right-hand side of Eq.~\ref{equadiff}.
The atomic diffusion velocity $v_i$ can be expressed as
\begin{equation}
v_i=D_{ip}\left[-\frac{\partial \ln X_i}{\partial r}+\frac{A_i m_p}{k T}(g_{rad,i}-g)+\frac{({\bar{Z}_i}+1)m_p g}{2 k T}+\kappa_T\frac{\partial \ln T}{\partial r}\right],
\label{eqvdiff}
\end{equation}
where $D_{ip}$ is the diffusion coefficient of element $i$ relative to protons; $g_{rad,i}$ the radiative acceleration on element $i$; $g$ the local gravity;  $\bar{Z}_i$ the average charge (in proton charge units) of element $i$ (roughly equal to the charge of the `dominant ion'); $k$ the Boltzmann constant; $T$ the local temperature; and $\kappa_T$ the thermal diffusivity.

The CESTAM code computes atomic diffusion, including radiative acceleration, taking into account partial ionisation for H, $^3$He, $^4$He, $^6$Li, $^7$Li, $^9$Be, $^{11}$B, $^{12}$C, $^{13}$C, $^{14}$N, $^{15}$N, $^{15}$O, $^{16}$O, $^{17}$O, $^{22}$Ne, $^{23}$Na, $^{24}$Mg, $^{27}$Al, $^{28}$Si, $^{31}$P (without radiative acceleration), $^{32}$S, $^{40}$Ca, and $^{56}$Fe. Radiative accelerations for light elements (below carbon) were not computed because they are negligible. Radiative accelerations were computed using the Single Valued Parameters (SVP) approximation \citep{alecian02,leblanc04}. The uncertainty of $g_{rad,i}$ provided by this method is about $30\%$ (Georges Alecian, private communication).

When rotation is included in the models, the turbulent diffusion coefficient is added to $D_{\rm turb}$. A description of the treatment of rotation in the CESTAM code can be found in \citet{marques13}. We considered the magnetised wind braking of \citet{matt15,matt19} and an additional vertical viscosity of $\nu_v=10^8$~cm$^2$s$^{-1}$ as calibrated by \citet{ouazzani19} to take into account the fact that the current rotation theory underestimates the transport of angular momentum. The other aspects of these models are the same as the ones presented in \citet{deal20}. These models are only used to show that rotation may explain Li depletion on the main sequence.

The implementation of turbulent mixing $D_\mathrm{turb}$ follows the standard phenomenological recipe \citep{richer00, richard01,richard02, michaud11}. The prescription for $D_\mathrm{turb}$ is not grounded in any ab-initio model, but is chosen not to affect the transport of the chemical elements close to the centre, as the efficiency of mixing drops with $\rho^{-n}$:
\begin{equation}\label{eqturb}
D_\mathrm{turb}=\omega D_{He}(T_0)~\left( \frac{\rho(T_0)}{\rho} \right)^n
,\end{equation}
where $\omega$ and $n$ are constants, $\rho$ is the density, and $D_{\rm He}(T_0)$ the atomic diffusion coefficient of helium at the reference temperature $T_0$. We assume $\omega=400$ and $n=3$. The atomic diffusion coefficient of helium was obtained using an analytical approximation as described by Eq.~\ref{eq:D_He}: 

\begin{equation}
\label{eq:D_He}
    D_{\rm He}=3.3 \times 10^{-15} T^{2.5} / 
    [ 4 \rho \ln(1 + 1.125 \times 10^{-16} T^{3} /\rho)],
\end{equation}
 \citep{michaud11} and it depends therefore on local conditions. As \citet{Richard2005} and \citet{Gruyters2016} showed, these values provide a good fit to the observed data of metal-poor clusters. 
The parameter $\rho(T_0)$ represents the density at the reference temperature. In this discussion, we used $T_0$ (see Eq.\ref{eqturb}) as a proxy of the turbulent mixing efficiency. By choosing different values of $T_0$, we effectively change the spatial extension of the zone, in which the turbulent mixing takes place. 
In other words, larger $T_0$ corresponds to a deeper base of the zone subject to the additional mixing and a larger turbulent diffusion coefficient. Therefore, the additional mixing is more efficient at competing with the atomic diffusion at higher  $T_0$ values.
\subsection{The impact of atomic diffusion and mixing on chemical abundances}\label{sec:s43}

Figure \ref{dDturbplot} shows the behaviour of the turbulent mixing coefficient $D_\mathrm{turb}$ in the models of a $1.0$ $\msun$ star (left panels) and $1.4$ $\msun$ star (right panels) with an age of $2.5$ Gyr and [Fe/H]$_\mathrm{ini}=-0.05$ dex. The convective core is located at $\log T > 7.145$ for the 1.4 $\msun$ model and is not shown in the plot. In the top panels, we colour-coded the profiles of $D_\mathrm{turb}$ by their corresponding $T_0$ values. The $T_0 = 10^{6.0}$ K model (solid red line) corresponds to the maximum penetration of the deep mixing zone, down to $R/R^* \sim 0.4$ in both models. On the other hand, the $\log T_0 = 10^{5.50}$ K model (blue line) has a very shallow mixing zone, which only reaches down to $R/R^* \sim 0.7$. The atomic diffusion coefficient of He is marked with the dashed line; this quantity represents a mean efficiency of atomic diffusion. Also, the profile of rotational mixing, $D_\mathrm{rot}$, with zero-age MS velocities of $\upsilon = 2.2$ and $7.0$ km/s, are indicated. 

The lower panel of Figure \ref{dDturbplot} shows the profiles of Li abundances corresponding to all aforementioned models. To aid the interpretation of this figure, we also show the surface convective zone of the model with $T_0 = 10^6$ K (grey area) and the Li destruction zone (T$>2.5\times10^6$~K, green area). The radial extent of the surface convective zone is slightly different for the models with other $T_0$ values. This figure helps us to understand why the surface abundance of Li is so sensitive to the exact prescription adopted for turbulent and rotational mixing. In particular, the surface abundance of Li decreases faster while either (a) decreasing the size of the turbulent mixing zone (decreasing $T_0$) or (b) increasing the depth of the convective envelope (or decreasing the mass of a star). In the former case, the larger the $T_0$, the stronger the mixing below the convective envelope, which counterbalances the effects of atomic diffusion (driven by gravitational settling) on Li. In the latter case, the model with a lower mass (1.0 $\msun$) has a deeper convective envelope that acts as a source of efficient mixing, and, therefore, quenches atomic diffusion. Turbulent mixing helps to avoid strong surface abundance variations of all chemical elements, but it is not efficient enough to bring Li down to its nuclear destruction region. On the contrary, rotation can induce a Li destruction but cannot reduce strong surface abundance variations of chemical elements \citep{deal20}. Indeed, the 1.0 $\msun$ model, which includes rotational mixing, reduces the surface Li abundance by more than one order of magnitude. This effect is much less important for more massive stars, owing to their shallower convective envelopes.  

\begin{figure}[]
\centering
\includegraphics[width=\linewidth]{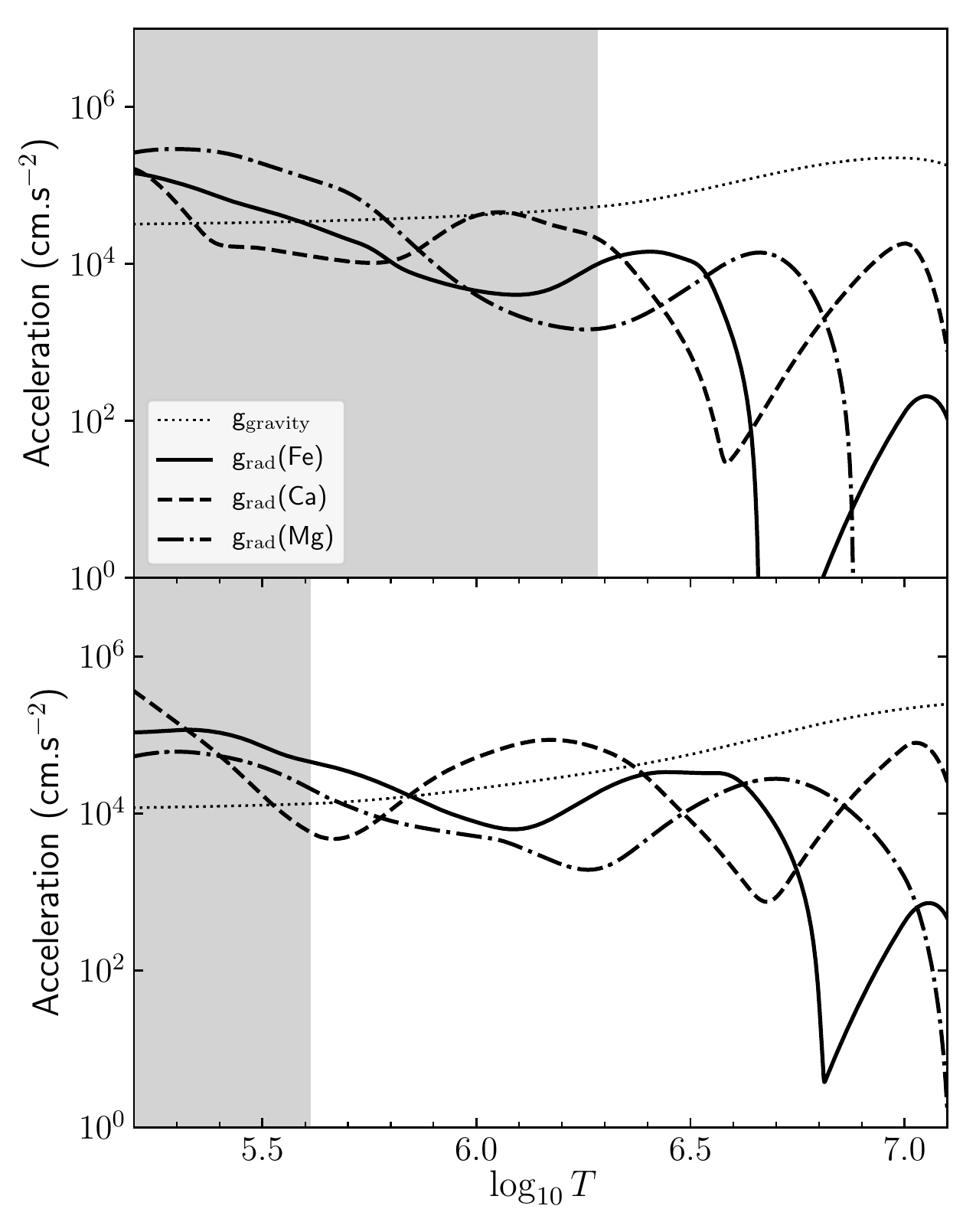}
\caption{Acceleration of gravity and radiative accelerations for Fe, Ca, and Mg according the the temperature of a 1.0 $\msun$ (upper panel) and 1.4 $\msun$ (lower panel) at 2.5~Gyr with an initial [Fe/H] of -0.05. The grey areas show the extension of the surface convective zones.}
\label{gradplot}
\end{figure}

Figure~\ref{gradplot} shows the profiles of radiative acceleration for Fe, Ca, and Mg, in comparison with the profile of gravitational acceleration for the same 1.0 $\msun$ (upper panel) and 1.4 $\msun$ (lower panel) models. By comparing the profiles of $g_{\rm rad}$ and $g_{\rm grav}$, we can see how the balance of two forces modifies the behaviour of the elemental abundances with depth. In general, owing to $g_{\rm rad} > g_{\rm grav}$ at the bottom of the surface convective zone, heavy elements tend to accumulate at the surface of a star (or inside if the accumulation occurs deeper: \citealt{richard01,theado09,deal16}), more so in the more massive model with $1.4$ $\msun$. Additionally, even if $g_{\rm rad} < g_{\rm grav}$, radiative acceleration will moderate the efficiency of gravitational settling. The behaviour at greater depths is very non-linear, which is caused by the complex dependence of the opacity on the ionisation state of the elements, and, consequently, on the density and temperature profiles in the interior \citep{richer98}. If we consider no additional transport processes in the 1.4 $\msun$ model, Fe and Mg should be accumulated at the surface and Ca would be depleted.

\begin{figure}[]
\centering
\includegraphics[width=\linewidth]{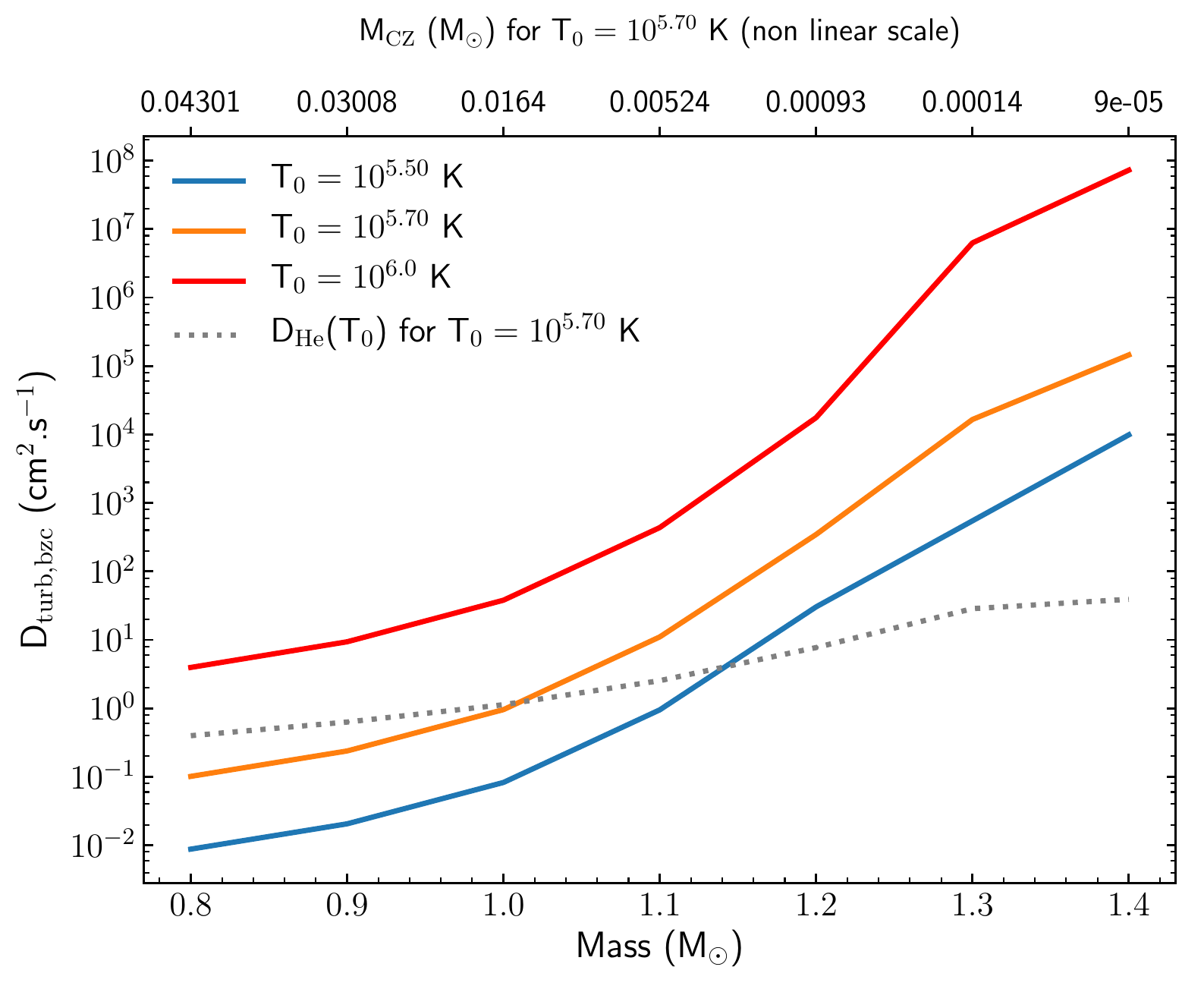}
\caption{$D_\mathrm{turb}$ for different values of $T_0$ at the bottom of the surface convective zone  according to the mass of the models. The masses of surface convective zones are shown in the upper x-axis.}
\label{dturbzc}
\end{figure}
\begin{figure*}
\centering
\includegraphics[]{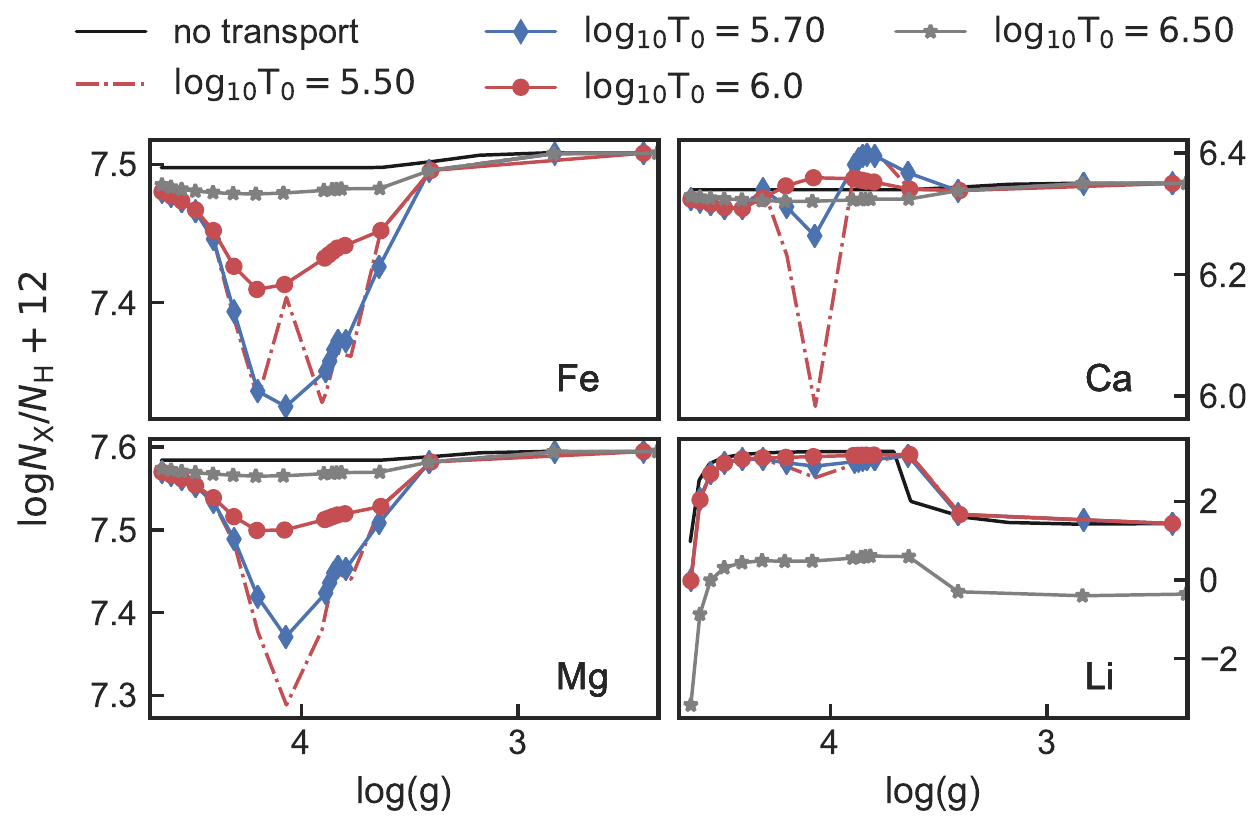}
\caption{Surface abundances against the $\log g$ as predicted by the models at $\mathrm{[Fe/H]_{\it initial}}=-0.05$. Different colours represent different transport prescriptions adopted in the models.}
\label{models_Dturb}%
\end{figure*}

Figure \ref{dturbzc} shows $D_\mathrm{turb}$ at the bottom of the surface convective zone for different $T_0$ (coloured solid lines) and, $D_\mathrm{He}(T_0)$ the helium diffusion coefficient at the temperature $T_0=10^{5.7}$~K (dashed line), according to the mass of stellar models. $D_\mathrm{turb}$ increases at the bottom of surface convective zone with increasing stellar mass for two reasons. The first one is due to the shallower convective zone in the more massive stars, implying that the reference temperature $T_0$ is far deeper from the bottom of the surface convective zone, which induces a large turbulent diffusion coefficient according to Eq.~\ref{eqturb}. The second reason is the $D_\mathrm{He}(T_0)$ increasing with the stellar mass as the internal structure in varying with mass at a given internal temperature, among other things $\rho$, which induces a $D_\mathrm{He}(T_0)$ value varying in two orders of magnitude between $0.8$ and $1.4$ $\msun$.

The model predictions for the surface abundances of Fe, Mg, Ca, and Li are shown in Fig.~\ref{models_Dturb}. All models correspond to the age of 2.5~Gyr, consistent with our observational constraints. Considering an age uncertainty of 0.2~Gyr leads to the difference in the predicted surface abundances of maximum 0.005~dex in the models for which atomic diffusion is the most efficient. The models differ in the value of $T_0$. Not surprisingly, the surface element abundances do not remain constant with the evolutionary phase of a star, unless extreme values of turbulent mixing, with $T_0$ superior to $10^6$~K  (equivalent to $D_\mathrm{turb}> 10^8$~cm$^2$.s$^{-1}$ at the bottom of the surface convective zone for the 1.4~M$_\odot$ model as shown in Fig.~\ref{dturbzc}), are adopted. Similarly, the effect of radiative levitation is to prevent the 'metal sink' effect of gravitational settling for some chemical elements, allowing the model to avoid critical surface under-abundances of metals. 

The evolution of Mg and Fe along the isochrone is very similar. The abundances of both elements are significantly depleted at the TO point of the cluster, which corresponds to $\log g \approx 4.3$ dex for the most metal-poor model and to $\log g \approx 4.1$ dex for the solar metallicity model. Ca, in contrast, displays a modest overabundance at the TO point, although this behaviour can be inverted for certain combinations of $\log T_0$ and initial abundances. Increasing $\log T_0$ generally flattens the tracks of [A/H] towards the initial abundances as the value of D$_{turb}$ increases at the bottom of the surface's convective zone (Fig. \ref{dturbzc}). The transition from the main sequence to the sub-giant phase is associated with a large depletion of Li, as the surface convective envelope deepens and Li-poor material is advected to the surface.

Of note is the non-linear behaviour of the models in the domain of inefficient turbulent mixing, $\log T_0 \leq 5.50$. In this regime, turbulent mixing is no longer sufficient to balance the outward radiation force. This imbalance induces a relative accumulation of an element at the surface, which may even result in an overabundance of the element relative to its initial unperturbed value. This process is responsible for the characteristic bump in the behaviour of  Ca abundances in the transition region between the TO point and the sub-giant branch and globally reduces the Fe and Mg depletion. This kind of effect can also be seen on Fe in Fig. \ref{models_Dturb}.
\begin{figure*}[ht!]
\centering
\includegraphics[scale=0.9]{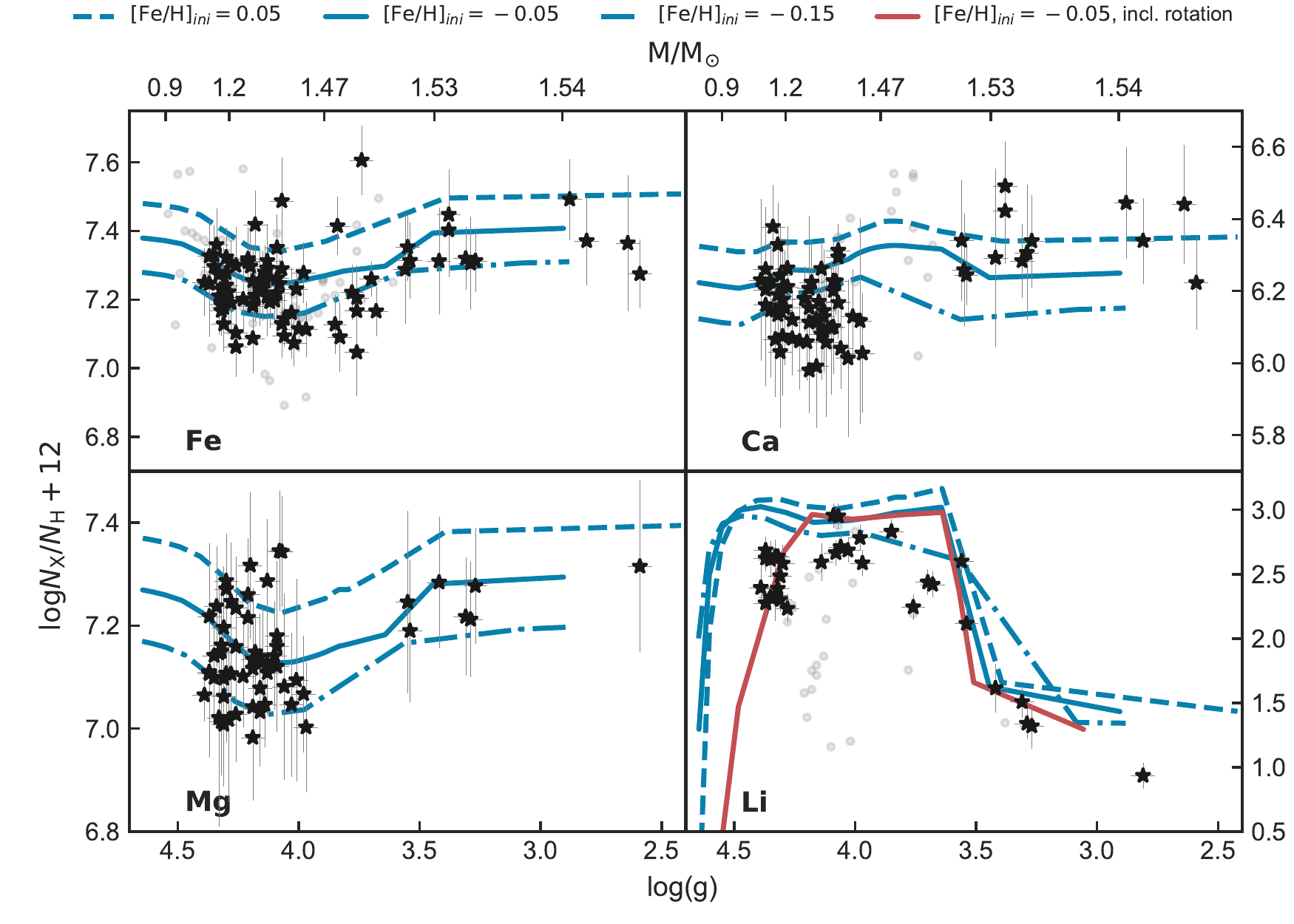}
\caption{3D NLTE photospheric abundances of Fe, Mg, Ca, and Li in stars of the NGC 2420 cluster (in black). Grey background symbols depict non-detections or abundances derived from spectra with low S/N and/or high rotational broadening. Lines correspond to the model predicted abundance variations due to atomic diffusion, incl. radiative acceleration, and turbulent mixing (T5.80 prescription) at various \textit{\emph{initial}} metallicities. The models are computed with solar chemical mixture for all elements except Mg, which shows an abundance at the RGB stage 0.2 dex lower than the Sun. The red line corresponds to the model with included rotation in addition to the T5.80 turbulent mixing that successfully explains observed Li abundance on the low main sequence. For more details on comparison of observations and models, see Sect. \ref{obs_vs_models}.}
\label{abundances_3D_NLTE}%
\end{figure*}
\section{Results}\label{results}
Over the past decade, several observational studies reported clear and systematic evolutionary trends in the chemical abundances in open and globular clusters. These trends are commonly attributed to the effect of atomic diffusion. However, it was also shown that atomic diffusion alone is not sufficient to explain the observations of different chemical elements. Therefore, additional mixing processes of a macroscopic nature were put forward \citep[e.g.][]{Mucciarelli2011, Korn2007,nordlander12,Gruyters2016}.

In this section, we compare our new observational data with different models described in Sect. \ref{models} and discuss the results in the context of other empirical and theoretical studies.
\subsection{Intra-cluster abundance variations}
Our 3D NLTE distributions of chemical abundances in \object{NGC~2420} stars are shown in Fig. \ref{abundances_3D_NLTE}. The measured abundances of Fe, Ca, and Mg are significantly lower at the cluster TO point, with a maximum depletion of $-0.2$ dex relative to the lower MS or RGB stars. This prominent under-abundance gradually disappears along the SGB, and the abundances attain their original (birth composition) values at the base of the RGB, around $\log g \approx 3.3$ dex. The depletion at the cluster TO is also predicted by stellar models computed with atomic diffusion and turbulent mixing. One such model, computed using the $T_0$ of $10^{5.8}$~K and several values of initial metallicity, is over-plotted onto the observed data. The undulations of theoretical profiles for stars in the mass range from 1.1 to 1.3 $\msun$ ($\logg$ range from 4.5 to 3.5) are caused by the interplay of gravitational settling, radiative pressure, and turbulent mixing, as described in Sect. \ref{sec:s43}. However, our data are not accurate enough to resolve these tiny signatures, which would require an abundance accuracy better than $0.05$ dex. Nonetheless, the global systematic trends in the data agree very closely with the models. On the RGB, the deepening of the surface convective zone after the MS stage quickly restores the surface abundances of elements, which are not affected by nuclear reactions (Fe, Ca, Mg), to their initial value.
\begin{figure*}[ht!]
\centering
\includegraphics[scale=0.9]{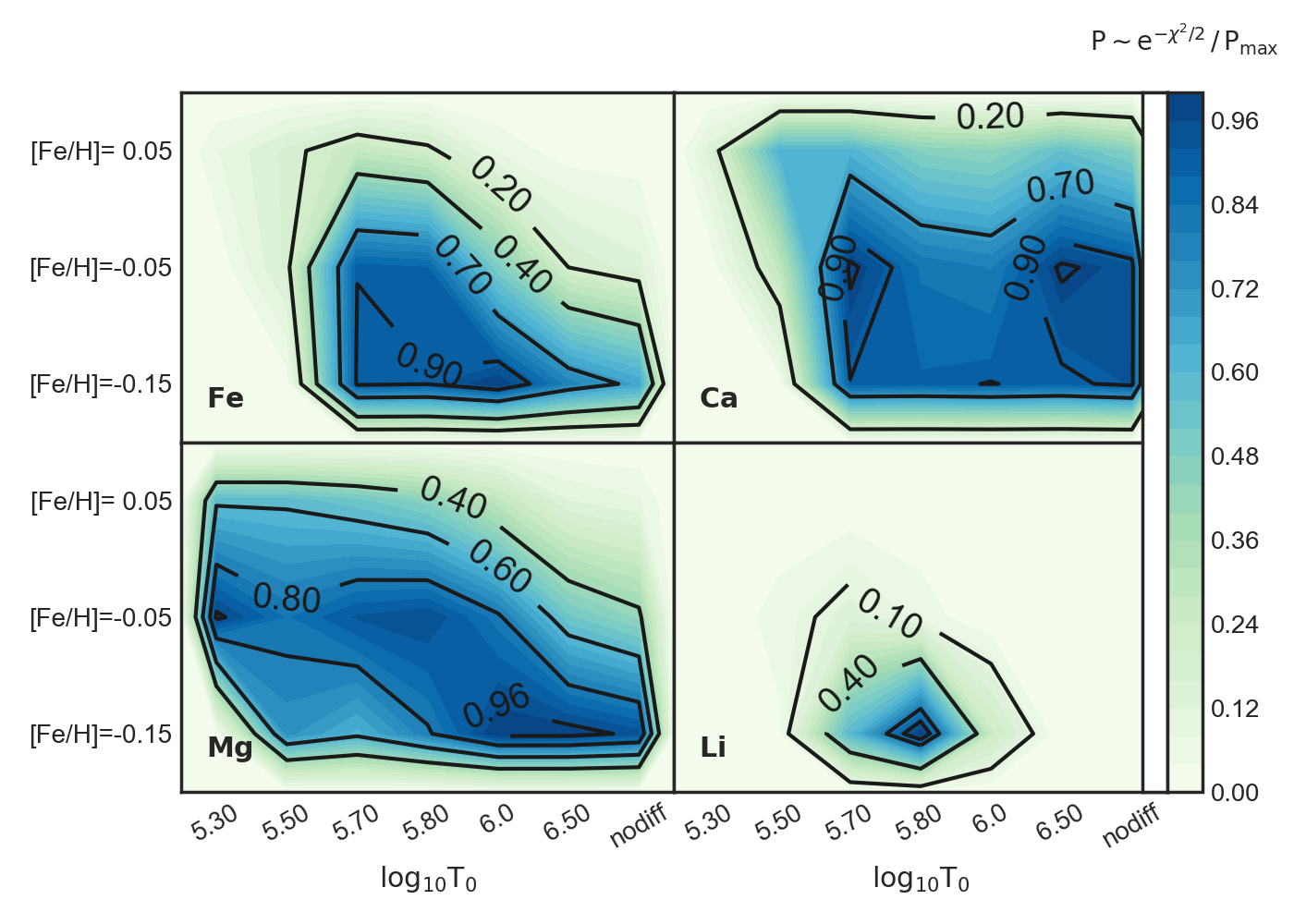}
\caption{Probability distributions for all stars in NGC 2420. We note that each panel represents one chemical element only (see inset).}
\label{likelihood_separate}%
\end{figure*}
\subsection{Distribution of Li abundances}

The behaviour of Li in the cluster is qualitatively different from the behaviour of other elements. In contrast to other elements, the profile of Li with $\logg$ cannot be fully described by our standard models that include atomic diffusion and turbulent mixing only. As seen in the bottom right panel of Fig. \ref{abundances_3D_NLTE}, the observed abundances of Li on the MS (at $\log g \sim$ 4.3 dex) are lower, compared to standard models. However, this problem can be solved by additionally including internal rotation, because this physical mechanism acts throughout the entire star and, therefore, allows for mixing in the deeper regions, which are hot enough to destroy Li. In the model computed with $\upsilon \approx$ 7.0~km/s, transport induced by internal rotation leads to earlier destruction of Li and results in the characteristic depletion of Li abundance on the MS, which is in agreement with the observations.  

In addition, at $\log g \sim$ 4.3 dex, there are a few data points  that suggest the presence of the Li dip \citep{Deliyannis2019}, that is the depletion of Li abundances compared to stars with higher and lower $\log g$ values. This dip was also seen in other light elements, like Be \citep{Smiljanic2010}, but its origin is still debated. 

The abrupt depletion of the Li abundance on the SGB, predicted by the model, is also seen in our observational data. The first dredge-up brings highly processed Li-poor material from the interior to the surface of a star. Consequently, the Li abundance drops by two orders of magnitude at the base of the RGB. The CESTAM models with rotation predict a lower abundance of Li on the SGB phase due to the deeper mixing induced by rotation.

The effect on the other elements is much smaller as rotation only impacts the size of the region where the chemical composition is homogenised. In the case of the $T_0 = 10^{5.8}$~K, the region is extended by only a few percent in mass, which leads to the change of observed abundance of $\sim$0.07~dex. In contrast to that, atomic diffusion and the parametrised mixing lead to the depletion of Fe by 0.2 dex. The impact of adding rotation on the other element will be investigated in a forthcoming paper.
\subsection{Combined statistical analysis of data and models} \label{obs_vs_models}
Figure~\ref{likelihood_separate} depicts the probability maps, which we employed to constrain the models consistent with our data. The maps were constructed by using a grid of CESTAM tracks for a cluster of a given age and different initial metallicities, as described in Sect.~\ref{models}. We performed the comparison between the observed abundance trends in \object{NGC~2420} and a grid of CESTAM models by computing a likelihood for each of these models as described by Eq.~\ref{eq:L_chi}, where $\chi^2_{k}$ is a chi-square per degree of freedom showing the goodness of fit by a certain model, and $\mathrm{[X/H]^{Obs/Mod}_{i}}$ are observed and theoretically predicted abundances, respectively. Statistical and systematic observational uncertainties described in Sect.~\ref{uncertainties} were taken into account. To ease the comparison between different elements, we present normalised values of the likelihoods on the maps. This representation does not affect our conclusions.
In the following,\begin{equation}
\label{eq:L_chi}
  P = e^{-\chi^2_{k} /2},\,\,\,
  \chi^2_{k} = \left( \sum^{N_{obs}}_{i}\frac{(\rm{[X/H]^{Obs}_{i}-[X/H]^{Mod}_{i})^2}}{\sigma_{\rm{[X/H]^{Obs}_{i}}}^2} \right) / k
,\end{equation}
each chemical element clearly shows a different sensitivity to the input parameter space of the models. 
For some elements, namely Ca, it is barely possible to distinguish stellar models with different levels of physical complexity.
One should also take into account that the response to transport processes is different for each element. In Sect.~\ref{sec:s43}, we show that the Li abundance on the MS is very sensitive to rotation, while for other elements the effect of rotation is negligibly small.
Thus, the abundance of Li alone would not be sufficient to identify the most likely model.
The degeneracy can be broken by combining the constraints from all four chemical elements simultaneously. 

\begin{figure}[ht!]
\centering
\includegraphics[scale=0.71]{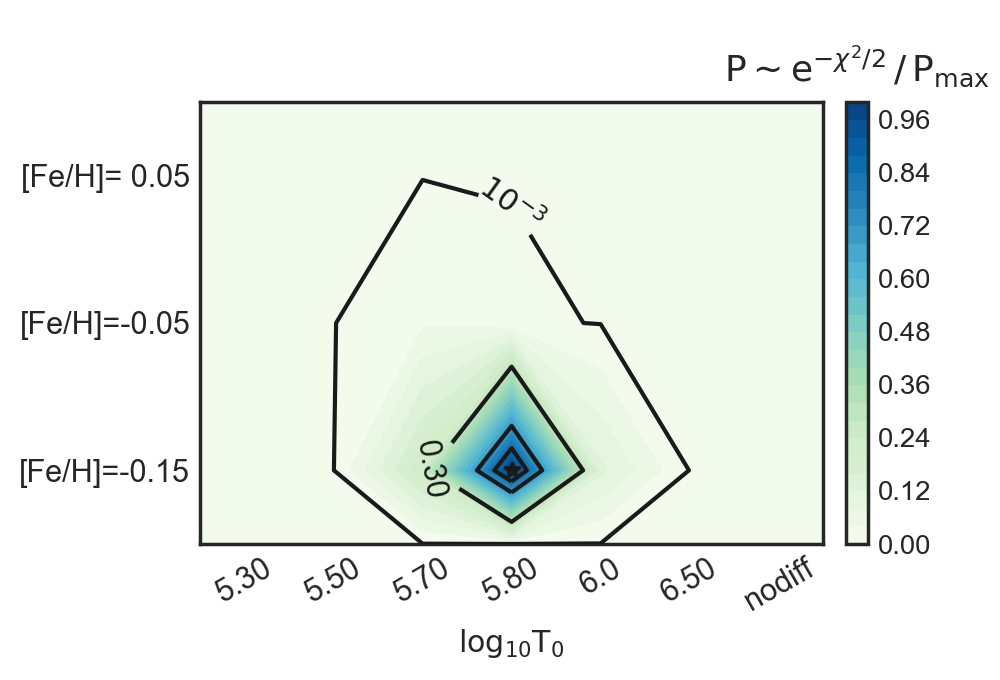}
\caption{Combined probability distributions for all stars in NGC 2420, computed using all four elements simultaneously. The most probable solution corresponds to the CESTAM model with atomic diffusion and a modest efficiency of turbulent mixing ($\log T_0 = 5.70$).}
\label{likelihood_combined}
\end{figure}

The combined analysis of all elements in our data set suggests that the model that is favoured by our data has a modest mixing efficiency of $\log T_0 = 5.8$ (Fig. \ref{likelihood_combined}). This model reproduces the intra-cluster distributions of the abundances of all measured chemical elements against fundamental parameters of stars. 
Other models computed assuming significantly higher ($\log T_0 = 6.0$) or significantly lower ($\log T_0 = 5.5$) efficiencies of turbulent mixing, or assuming no atomic diffusion at all, are not supported by our data.

In summary, our observational data for \object{NGC 2420} yields strong constraints on the physical processes in stellar interiors. Combining four chemical elements - Fe, Ca, Mg, and Li - we are able to confine the parameter space of the stellar models and to identify the most probable range of values that characterise the efficiency of turbulent mixing at the bottom of the stellar convective envelope. The CESTAM model, which includes atomic diffusion, turbulent mixing with $\log T_0 = 5.8$, and rotational mixing, is best supported by our data. Other stellar models, in particular including those computed without atomic diffusion, are ruled out with a high level of confidence.

\section{Discussion}\label{discussion}

Similar attempts to constrain the transport of chemical elements in the stellar interior using observations were carried out in several other studies \citep[e.g.][]{Korn2006, Korn2007, Mucciarelli2011, nordlander12, Gruyters2014, Gruyters2016}. However, most of these studies are limited to the analysis of old (over 10 Gyr) and metal-poor ([Fe/H] $\lesssim -1$ dex) Galactic clusters. Using the globular clusters NGC 6397 ([Fe/H] $= -2$ dex) and M 30 ([Fe/H] $= -2.3$ dex) they constrained $\log T_0 = 6$. This value is at the upper boundary of our estimated range of $\log \mathrm{T_0}$ from $5.70$ to $6.00$. The difference with our results, which are based on the young cluster with a slightly sub-solar metallicity, \object{NGC 2420}, may indicate a metallicity dependence of the efficiency of mixing processes, such as rotational and turbulent mixing, competing with atomic diffusion. This metallicity dependence is clearly present in the stellar evolution models (see Fig. 8: compare models computed with [Fe/H] $=$ -0.15 and $0.05$ dex) and combining the results from the aforementioned studies, we can conclude that there is significant evidence for a different efficiency of T$_0$ at lower metallicities.

Comparing our results with previous studies of open clusters, we find a good agreement in the slopes of abundance trends and behaviour of individual chemical elements. M67 - a 4 Gyr old solar-metallicity cluster - is arguably the best studied system in this respect \citep{Yong2005, Randich2006, Oenehag2014, Motta2018, Souto2018, Gao2018}.
The last study is similar to our work in that it is based on the NLTE analysis of the abundances of several chemical elements. They use the MESA stellar models from \citet{Dotter2017} computed with account for atomic diffusion, radiative acceleration, overshoot, and turbulent mixing calibrated on the observed data for NGC 6397 from \citet{Korn2007}. However, although their findings are generally consistent with the MESA models, they stress non-negligible differences on the RGB and red clump, especially for the key elements, such as Mg, Na, and Fe. This conclusion is corroborated by \citet{Souto2019}, who also included odd-Z species such as N, K, V, and Mn, in the analysis. They confirm systematic depletion of metals at the cluster TO, as seen in previous studies. They also find a non-negligible abundance spread at all evolutionary phases. The most recent study of M67 by \citet{Liu2019} based on 3D NLTE and 1D NLTE for selected chemical elements confirms chemical inhomogeneity in M67, reinforcing evidence for signatures of atomic diffusion in the cluster. Whereas they make no attempt to quantitatively constrain the mixing processes, they emphasise the need to include such models in studies of stellar populations and chemical evolution. 

\citet{verma19} showed that the analysis of acoustic glitches in asteroseismic data can be used to constrain turbulent mixing. Their results for three Kepler targets yielded $\log \mathrm{T_0}$ in the range from $5.9$ to $6.0$, when comparing the effect on surface abundances, which is also slightly higher compared to our results. We note, however, that their parametrisation of turbulent mixing (and hence, their definition of $T_0$) is not exactly the same as ours.  Nevertheless, the conclusions of our and their study are mostly similar, taking into account both models and observations uncertainties. This fact is reassuring because the two methods to constrain the efficiency of mixing processes in stellar interiors are entirely independent.
\section{Stellar structure and evolution}\label{perspectives}
Our findings concerning the relevance of atomic diffusion and mixing in stellar evolution are important in the context of other areas of astrophysics. 

The most obvious consequence of our study is that accurate identification of membership to stellar associations and open and globular clusters \textit{\emph{cannot}} rely on metallicity. This is still a common procedure in studies of stellar clusters \citep[e.g.][]{Blanco-Cuaresma2018, Donor2020}. However, it obviously leads to biases in the population statistics and determination of the age of cluster and its metallicity.

The next interesting consequence arises in the context of chemodynamical structure and evolution of the Galaxy.
It is common to associate the present-day position of a star and its observed abundance pattern with its initial chemical composition, corresponding to that of the interstellar matter or star-forming region in which the star formed \citep{Casagrande2011, Bensby2014, Bergemann2014, Recio-Blanco2014, Hansen2019, Hayden2020}, except for the effects of kinematic mixing and radial migration in the Galaxy \citep[e.g.][]{Schoenrich2009}.
This information is used to infer quantities describing the present-day Galactic structure, ignoring the significant systematic effects that secular stellar evolution has on the photospheric abundances of stars. In turn, this causes a systematic bias in radial gradients, metallicity distribution functions, and even age-metallicity relationships, because the determinations of stellar ages by means of isochrone fitting are also affected by the problem of selective enhancement or depletion of abundances in different evolutionary phases \citep[e.g.][]{JofreWeiss2011, Salaris2016, Dotter2017}.

Our results suggest that stellar abundances can be used to constrain the history of the Galaxy under the condition that stellar evolution models, and therefore, the stellar yields, implemented in Galactic chemical evolution (GCE) models, take atomic diffusion and mixing into account. An alternative solution is to restrict the analysis to the samples of stars, for which observed abundance patterns are not significantly affected by diffusion and mixing. The available evidence suggests that RGB, young ($<1$ Gyr) ,and slowly rotating ($<5$ km/s) stars, as well as low-mass ($<0.9 \msun$) unevolved main-sequence stars, are relatively robust tracers of the initial composition of the interstellar matter, whereas solar-like main-sequence stars, turn-off stars, sub-giants (which dominate local Galactic neighbourhood and are very populous in samples like the GALAH \citep{Buder2019}) are to be treated with caution. Moreover, this selective approach requires accurate determination of stellar masses and robust statistical modelling of selection functions in order to quantify the bias in the population statistics arising from target selection.
\section{Conclusions}\label{conclusions}
We presented a homogeneous analysis of Gaia-ESO spectra of stars in the open cluster \object{NGC 2420} - a relatively young (2.5 $\pm 0.5$ Gyr) and metal-rich massive cluster at a distance of $\sim 2.5$ kpc \citep{Cantat-Gaudin2018}. About $30 \%$ of stars in the cluster could be unresolved main-sequence binaries according to the method described in \citet{Cordoni2018}.

The spectra were taken with the Giraffe medium-resolution ($19\,200 \leq R \leq 21\,500$) spectrograph at the VLT. We combined our spectroscopic analysis with photometry and astrometry from Gaia DR2. Our sample includes $\sim 84$ stars and covers the full evolutionary sequence in the cluster, from G-type stars on the main sequence to K-type red giants. We used NLTE atomic models, as well as 1D hydrostatic (MARCS) and averaged 3D hydrodynamical (STAGGER) model atmospheres, to determine the abundances of Fe, Ca, and Mg in the cluster stars. The abundances of Li were measured using 1D LTE models and corrected for 3D NLTE effects using literature values. 

We find that the chemical abundance distributions in the cluster display significant trends with the evolutionary stages of the stars. Furthermore, Fe, Mg, and Ca show a $\sim 0.1$ to $0.2$ dex depletion at the cluster TO point, but the abundances gradually increase and flatten near the base of the RGB. The abundances of Li are low for stars with $M \lesssim 1 \msun$, but increase for higher mass stars and remain relatively constant at the level of A(Li) $=2.8$ dex at the cluster TO. This value is close to the value predicted by the the standard models of big bang nucleosynthesis (SBBN). The Li abundances drop by two orders of magnitude on the SGB, attaining A(Li) $=1.3$ dex on the RGB. 

We attribute the systematic difference in abundances in the cluster to atomic diffusion and mixing. Comparing our findings with CESTAM stellar evolution models \citep{deal18}, we find that only RGB stars with masses $\gtrapprox 1.5 \msun$ ($\logg \lesssim 3.5$ dex) can be viewed as robust tracers of the initial composition of the cluster. Also, low-mass stars with $M \lessapprox 0.9 \msun$ are not expected to display self-processed photospheric abundances. 
Therefore, the initial chemical composition of \object{NGC 2420} is A(Fe) $= 7.35 \pm 0.1$ dex, A(Mg) $= 7.3 \pm 0.1 $ dex, A(Ca) $= 6.4 \pm 0.1$ dex, A(Li) $= 2.8 \pm 0.1$ dex. The present-day composition at the cluster turn-off is significantly different: A(Fe) $= 7.15 \pm 0.1$ dex, A(Mg) $= 7.15 \pm 0.1$ dex, A(Ca) $= 6.14 \pm 0.1$ dex. We emphasise that these chemical offsets between the low-mass and higher mass stars are caused by physical processes during stellar evolution, and are, consequently, essential to take into account in any study that uses stellar abundances for detailed diagnostics of stellar structure, exoplanet characterisation, or Galaxy history and formation.

The results obtained in 1D LTE, 1D NLTE, and in 3D NLTE show systematic differences. The RGB stars are more sensitive to NLTE effects, compared to main-sequence and turn-off stars. Our estimates of the differences between 3D NLTE and 1D LTE abundances for RGB stars amount to $\delta$(Fe) $= 0.06$ dex, $\delta$(Mg) $= 0.01$ dex, and $\delta$(Ca) $= 0.04$ dex. For the TO stars, the 3D NLTE effects are within $0.1$ dex for all elements.

We carried out a systematic, statistical analysis of the observed abundance distributions using the grid of CESTAM models, computed with atomic diffusion and different prescriptions for turbulent mixing, with and without mixing induced by rotation, as well as different initial metallicities. 
The combined probabilistic analysis allows us to confine the parameter space of the models and to constrain the depth of the zone in which turbulent mixing takes place, providing in the sense a quantitative assessment of the combined efficiency of microscopic and macroscopic mixing. 
We find that the most likely model, which is consistent with the observed trends of all elements, has a $\log T_0$ of 5.7 to 6.0. This value is slightly lower than the previous literature estimates, which are based on the analysis of abundances in old globular clusters and acoustic glitches in astroseismic data, although we warn that detailed comparisons in the latter case are hampered by different formulations of the turbulent coefficient.

\begin{acknowledgements}
The work of E.S. was partially funded by the subsidy 3.9780.2017/8.9 allocated to Kazan Federal University for the state assignment in the sphere of scientific activities.
MD acknowledges support by FCT/MCTES through national funds (PIDDAC) by these grants UIDB/04434/2020, UIDP/04434/2020 and PTDC/FIS-AST/30389/2017 and by FEDER - Fundo Europeu de Desenvolvimento Regional through COMPETE2020- Programa Operacional Competitividade e Internacionaliza\c c\~ao by this grant POCI-01-0145-FEDER-030389. MD is supported in the form of a work contract funded by national funds through Funda\c{c}\~{a}o para a Ci\^{e}ncia e Tecnologia (FCT). M.D. acknowledges financial support from the "Programme National de Physique Stellaire" (PNPS) of the CNRS/INSU co-funded by the CEA and the CNES, France.
We acknowledge support by the Collaborative Research centre SFB 881 (projects A5, A10), Heidelberg University, of the Deutsche Forschungsgemeinschaft (DFG, German Research Foundation). 
A.S.~is partially supported by the grants ESP2017-82674-R (Spanish Government) and 2017-SGR-1131 (Generalitat de Catalunya).  
TB was funded by the project grant "The New Milky Way" from the Knut and Alica Wallenberg Foundation,
and project grant No. 2018-04857 from the Swedish Research Council.
This work has made use of data from the European Space Agency (ESA) mission
{\it Gaia} (\url{https://www.cosmos.esa.int/gaia}), processed by the {\it Gaia}
Data Processing and Analysis Consortium (DPAC,
\url{https://www.cosmos.esa.int/web/gaia/dpac/consortium}). Funding for the DPAC
has been provided by national institutions, in particular the institutions
participating in the {\it Gaia} Multilateral Agreement.
Based on data products from observations made with ESO Telescopes at the La Silla Paranal Observatory under programme ID 188.B-3002. These data products have been processed by the Cambridge Astronomy Survey Unit (CASU) at the Institute of Astronomy, University of Cambridge, and by the FLAMES/UVES reduction team at INAF/Osservatorio Astrofisico di Arcetri. These data have been obtained from the Gaia-ESO Survey Data Archive, prepared and hosted by the Wide Field Astronomy Unit, Institute for Astronomy, University of Edinburgh, which is funded by the UK Science and Technology Facilities Council.
This work was partly supported by the European Union FP7 programme through ERC grant number 320360 and by the Leverhulme Trust through grant RPG-2012-541. We acknowledge the support from INAF and Ministero dell' Istruzione, dell' Universit\`a' e della Ricerca (MIUR) in the form of the grant "Premiale VLT 2012". The results presented here benefit from discussions held during the Gaia-ESO workshops and conferences supported by the ESF (European Science Foundation) through the GREAT Research Network Programme.
We thank Jan Rybizki for a valuable input concerning the data representation. We thank Andreas Korn for a discussion on modelling the transport of elements. We thank an anonymous referee for their comments and suggestions.
\end{acknowledgements}

\newpage
\bibliographystyle{aa}
\bibliography{lit}

\end{document}